\documentclass[journal,twoclumn]{IEEEtran}

\usepackage{multicol}
\usepackage{etoolbox}
\makeatletter
\patchcmd{\@makecaption}
  {\scshape}
  {}
  {}
  {}
\makeatletter
\patchcmd{\@makecaption}
  {\\}
  {.\ }
  {}
  {}
\makeatother

\usepackage{amsfonts}
\usepackage{mathrsfs}
\usepackage{mathtools}
\usepackage{amsfonts}
\usepackage{amssymb}
\usepackage{graphicx}
\usepackage{epsfig}
\usepackage{psfrag}
\usepackage{amsmath}
\usepackage{array}
\usepackage{cases}
\usepackage{eufrak}
\usepackage{subfigure}
\usepackage{cite,graphicx,amsmath,amssymb,color}
\usepackage{algorithmic}
\usepackage{algorithm}
\usepackage{subfigure}
\usepackage{bm}
\usepackage{multirow}
\usepackage{threeparttable}
\usepackage{array}
\usepackage{makecell}

\DeclareMathOperator*{\argmax}{argmax}

\newtheorem{Cla}{Claim}
\newtheorem{Thm}{Theorem}

\newtheorem{Prob}{Problem}

\newtheorem{Proof}{Proof}

\IEEEoverridecommandlockouts

\begin{document}

\setcounter{page}{1}
\title{Power-Efficient Wireless Streaming of Multi-Quality Tiled 360 VR Video in MIMO-OFDMA Systems}
\author{\IEEEauthorblockN{Chengjun Guo, Lingzhi Zhao, Ying Cui,~\IEEEmembership{Member,~IEEE},  Zhi Liu,~\IEEEmembership{Senior Member,~IEEE}, and Derrick Wing Kwan Ng,~\IEEEmembership{Fellow,~IEEE}}
\thanks{
C. Guo, L. Zhao and Y. Cui are with the Department of Electronic Engineering, Shanghai Jiao Tong University, Shanghai 200240, China.
Z. Liu is with
Graduate School of Informatics and Engineering, the University of Electro-Communications, Tokyo 182-8585, Japan.
D. W. K. Ng  is with the School of Electrical Engineering and Telecommunications,
University of New South Wales, Sydney, NSW 2052, Australia. This paper will be presented in part at the IEEE ICC 2021 \cite{vr-icc21}.
}}

\maketitle
\thispagestyle{headings}


\begin{abstract}
In this paper, we study the optimal wireless streaming of a multi-quality tiled 360 virtual reality (VR) video
from a multi-antenna server to multiple single-antenna users in a multiple-input multiple-output (MIMO)-orthogonal frequency division multiple access (OFDMA) system.
In the scenario without user transcoding,
we jointly optimize beamforming and subcarrier, transmission power, and rate allocation to minimize the total transmission power.
This problem is a challenging mixed
discrete-continuous optimization problem.
We obtain a globally optimal solution for small multicast groups, an asymptotically optimal solution for a large antenna array, and a suboptimal solution for the general case.
In the scenario with user transcoding, we jointly optimize the quality level selection,
 beamforming, and subcarrier, transmission power, and rate allocation to minimize the weighted sum of the average total
transmission power and the transcoding power. This problem is a two-timescale mixed discrete-continuous optimization
problem, which is even more challenging than the problem for the scenario without user transcoding.
We obtain a globally optimal solution for small multicast groups, an asymptotically optimal solution for a large antenna array, and a low-complexity suboptimal solution for the general case.
Finally, numerical results demonstrate the significant gains of proposed solutions over the existing solutions.
\end{abstract}

\section{Introduction}
\IEEEPARstart{A} 360 virtual reality (VR) video can be generated by capturing a scene of interest in all directions simultaneously with an omnidirectional camera  \cite{deterfov}.
It is predicted that the VR market will
reach 87.97 billion USD by 2025 \cite{market}.
Transmitting a 360 VR video over wireless networks enables users to experience immersive environments  without
geographical or behavioral restrictions.
As a 360 VR video has a much larger file size  than a traditional video, streaming an entire 360 VR video brings a heavy burden to wireless networks\cite{JSAC19,TCOM18,TCOM19}.
When watching a 360 VR video, a user perceives it from only one viewing direction  at any time, which corresponds to one part of the 360 VR video, known as field-of-view (FoV).
The tiling technique is widely adopted to improve the transmission efficiency for 360 VR videos \cite{tiling1}. Specifically, a 360 VR video is divided into smaller rectangular segments of the same size, known as tiles. Transmitting the set of tiles covering each predicted FoV can reduce the required communication resources without reducing the  quality of experience  \cite{deterfov,tiling1}. In practice, users may have heterogeneous conditions (e.g., channel conditions, display resolutions of devices, etc.). Pre-encoding each tile into multiple representations with different quality levels allows bitrate (quality)
adaptation according to a user's condition. Therefore, wireless streaming of a multi-quality tiled 360 VR video has received growing interest.

Recently, \cite{ACMMM18,ICC18,JESCS19} study optimal wireless streaming of a multi-quality tiled 360 VR video in single-user networks. Specifically, \cite{ACMMM18,ICC18,JESCS19} optimize the quality level selection \cite{ACMMM18,JESCS19} or transmission rate \cite{ICC18} to minimize the total distortion \cite{ACMMM18,ICC18}, or total utility \cite{JESCS19}. The proposed solutions for single-user networks in \cite{ACMMM18,ICC18,JESCS19} are not applicable in multi-user networks, as optimal resource sharing among users with heterogeneous channel conditions is not considered. In several VR applications, such as VR gaming,  VR military training, and VR sports \cite{TCOM20,Arxiv20}, a 360 VR video has to be transmitted to multiple users simultaneously. When a tile is required by
multiple
users concurrently,
multicast opportunities can be utilized to improve transmission efficiency.
In \cite{guo_tdma,guo_ofdma,multi-user1,multi-user2,multi-user3,long,long2}, the authors study the optimal wireless streaming of a multi-quality tiled 360 VR video to multiple users
by exploiting multicast
opportunities.
In particular, in our previous works \cite{guo_tdma,guo_ofdma},
we optimize transmission resource allocation to minimize the average transmission power for given video quality requirements of all users and optimize the encoding rate of each tile to maximize the received video quality for a given transmission power budget.
In \cite{multi-user1,multi-user2,multi-user3,long,long2}, the~authors
focus on optimizing
quality level selection for each tile.
 Specifically, in \cite{multi-user1,multi-user2,multi-user3}, the authors
 maximize the total utility of all users \cite{multi-user1,multi-user2} or minimize the distortion of video scenes \cite{multi-user3}, without considering any constraints on
quality variation.
Consequently, more multicast opportunities can be exploited to further improve  transmission efficiency.
Nevertheless, the obtained quality levels of adjacent tiles
may vary significantly, leading to poor viewing experiences \cite{multi-user1,multi-user2,multi-user3}.
In \cite{long}, the authors impose some constraints on quality variation while maximizing the total utility of all users to address this issue.
Although the restrictions on quality variation in \cite{long}
can alleviate quality variation in an FoV to a certain extent, they cannot guarantee  quality smoothness and are less mathematically tractable.
In \cite{long2},
user transcoding is adopted to ensure that
the quality levels of all received tiles in each FoV are identical when maximizing the total utility of all users.

Despite the fruitful research in the literature, the performance of wireless streaming of a tiled 360 VR video is still unsatisfactory. The results in
\cite{guo_tdma,guo_ofdma,multi-user1,multi-user2,multi-user3,long,long2} are all for single-antenna
servers, which cannot exploit spatial degrees of freedom for effective resource utilization.
The performance of wireless systems can be improved by deploying multiple antennas at a server  and designing efficient beamformers.
Among various multi-antenna technologies, MIMO-OFDMA is the dominant air interface for  5G broadband wireless communications,
as it can provide more reliable communications at high speeds.
For instance,
in
\cite{mo1,mo2,mo3}, the authors consider single-group multicast \cite{mo1}
and multi-group multicast \cite{mo2,mo3}.
Specifically,
in \cite{mo1}, the authors  consider the optimization of beamforming and power allocation to maximize the minimum user data rate.
In \cite{mo2}, the authors consider the subcarrier allocation and power allocation to maximize the sum rate.
However, the solutions proposed in \cite{mo1,mo2} are heuristic and hence have no performance guarantee.
\cite{mo3} studies the optimization of beamforming to minimize the total transmission power. A
stationary point of the beamforming design problem is obtained based on successive convex approximation. Note that in \cite{mo3}, messages on each subcarrier are associated with different beamforming vectors,  resulting
in a substantial increase in the number of variables and the computational complexity for
solving the optimization problem.

This paper considers the optimal wireless streaming of a multi-quality tiled 360 VR video from one server to multiple users in a MIMO-OFDMA system in the scenarios without and with user transcoding. With more advanced physical layer techniques than those in \cite{guo_tdma,guo_ofdma,multi-user1,multi-user3,long,long2}, we expect the stringent requirements for 360 VR video transmission to be better satisfied.  Assume that the set of tiles to be transmitted to each user has been determined and
each user's quality requirement
 is given.
The main contributions of this paper are summarized below.
\begin{itemize}
\item In the scenario without user transcoding, we exploit natural multicast opportunities and formulate the minimization of the total transmission power with respect to (w.r.t.) beamforming, subcarrier allocation, transmission power, and rate allocation as a multi-group multicast problem in the MIMO-OFDMA system.
This problem is a challenging mixed discrete-continuous optimization problem.
We obtain its optimal solutions for two special cases, namely, the case of small multicast groups (for different sets of tiles) and the case of a large antenna array, exploiting decomposition, continuous relaxation, and Karush-Kuhn-Tucker (KKT) conditions. We also obtain a suboptimal solution for the general case using continuous relaxation and difference of convex (DC) programming.
Note that previous works studying multi-group multicast in MIMO-OFDMA systems do not investigate special cases in which optimal solutions can be obtained \cite{mo2,mo3}.
Besides, the proposed multi-group multicast formulation with one beamforming vector for each subcarrier can achieve the same performance as the multi-group multicast formulation in \cite{mo3} which has one beamforming vector for each user and each subcarrier, but yields a substantially lower computational complexity for the general case.
\item In the scenario with user transcoding,
a flexible tradeoff between computation and communications
resource consumptions can be struck via exploiting transcoding-enabled multicast opportunities.
 We utilize  natural multicast opportunities and transcoding-enabled multicast opportunities and  minimize the weighted sum of the average total transmission power and the transcoding power by optimizing the quality level selection, beamforming, and subcarrier, transmission power, and rate allocation.
This problem extends the multi-group multicast optimization in the scenario without user transcoding, and it is a more challenging two-timescale mixed optimization problem.
For two special cases, we obtain the corresponding optimal solutions.
For the general case, we obtain a low-complexity suboptimal solution.
Note that the formulations in \cite{mo2} and \cite{mo3}, which consider only natural multicast opportunities, cannot provide an effective design in the scenario with user transcoding.

\item Finally, numerical results show substantial gains achieved by the proposed
solutions over existing schemes and  demonstrate the advantages of multicast and user transcoding in wireless streaming of a multi-quality tiled 360 VR video. Furthermore, numerical results illustrate that  the proposed low-complexity optimal solution obtained for  a large antenna array can achieve promising  performance when the number of antennas is moderate, demonstrating its effectiveness. 
\end{itemize}

Note that this work extends our previous results on wireless streaming of a multi-quality tiled 360 VR video in time division multiple access (TDMA) systems  \cite{guo_tdma,long,long2} and OFDMA systems \cite{guo_ofdma}.
The extensions are highly nontrivial due to the non-convexity w.r.t beamforming vectors. To the best of our knowledge, this is the first work providing an optimization-based design for wireless streaming of a multi-quality tiled 360 VR video in MIMO-OFDMA systems.

\emph{\bf Notation}: 
 For a Hermitian matrix $\mathbf{A}$, $\mathbf{A}\succeq\mathbf{0}$ means that $\mathbf{A}$ is an Hermitian positive semidefinite matrix. The symbol $(\cdot)^{H}$ denotes  complex conjugate transpose operator. $\text{tr}(\cdot)$ and $\text{rank}(\cdot)$ denote the trace and the rank, respectively.
 $\mathbb{E}[\cdot]$ denotes the statistical expectation.
 $\mathcal{CN}(\mathbf{a},\mathbf{R})$ represents the distribution of circularly-symmetric complex Gaussian  random vectors with mean vector $\mathbf{a}$ and covariance matrix $\mathbf{R}$.
 $\mathbf{I}_{N}$ denotes the $N\times N$ identity matrix.  $\mathbb{C}^{N\times M}$ denotes the space of $N\times M$ matrices with complex entries.

\section{System Model}
\begin{figure}[t]
\begin{center}
\subfigure[\small{Multi-quality tiled 360 VR video required by multiple users. }] 
 {\resizebox{7cm}{!}{\includegraphics{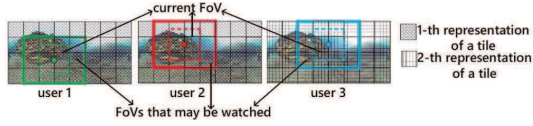}}}
 \subfigure[\small{Scenario without user transcoding.}]
 {\resizebox{7cm}{!}{\includegraphics{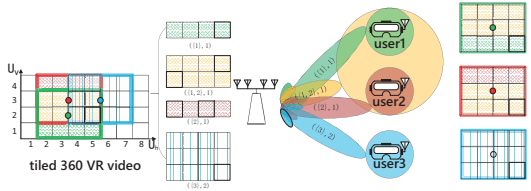}}}
  \subfigure[\small{Scenario with user transcoding.}]
 {\resizebox{7cm}{!}{\includegraphics{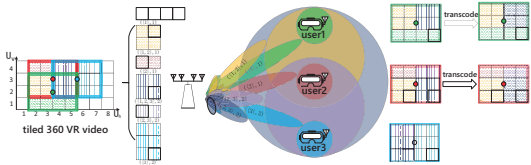}}}
 \end{center}
     \caption{\small{System models of wireless streaming of a multi-quality tiled 360 VR video in  the two scenarios. $K=3$, $\mathbf{r}=(1,1,2),$  $U_h\times U_v=8\times4$, $M=4$,
     $\mathcal{G}_1=\{(2,1),(3,1),(4,1),(5,1),(2,2),(3,2),(4,2),(5,2),(2,3),(3,3),$
     $(4,3),(5,3)\}$,
$\mathcal{G}_2=\{(2,2),(3,2),(4,2),(5,2),(2,3),(3,3),(4,3)$,
$(5,3),(2,4),$
$(3,4),(4,4),(5,4)\}$, $\mathcal{G}_3=\{(4,2),(5,2),(6,2),(7,2),(4,3),(5,3),$ 
$(4,4),(5,4),(6,4),(7,4)\}$,
 $\mathcal{I}=\{\{1\},\{2\},\{3\},\{1,3\},\{2,3\},\{1,2,3\}\}$,
 $\mathcal{P}_{\{1\}}=\{(2,1),(3,1),(4,1),(5,1)\}$,
 $\mathcal{P}_{\{2\}}=\{(2,4),(3,4)\}$,
 $\mathcal{P}_{\{3\}}=\{(6,2),(6,3),(6,4),(7,2),(7,3),(7,4)\}$,
 $\mathcal{P}_{\{1,2\}}=\{(2,2),(2,3),(3,2),(3,3)\}$,
 $\mathcal{P}_{\{2,3\}}=\{(4,4),(5,4)\}$,
 $\mathcal{P}_{\{1,2,3\}}=\{(4,2),(4,3),(5,2),(5,3)\}$,
 $\mathcal{K}_{\{1\},1}=\{1\},$
 $\mathcal{K}_{\{1,2\},1}=\{1,2\},$
 $\mathcal{K}_{\{2\},1}=\{2\},$
 $\mathcal{K}_{\{3\},2}=\{3\}.$
 }}
\label{fig:system-model}
\end{figure}

Wireless streaming of a multi-quality tiled 360 VR video from a server (e.g., access point or base station) to multiple users arises in several VR applications such as VR gaming, VR concert, VR military training, and VR sports.
This paper aims to optimize the wireless streaming of a multi-quality tiled 360 VR video from a server to $K$ users in a MIMO-OFDMA system as illustrated in Fig.~\ref{fig:system-model}.\footnote{We adopt a  multi-quality tiled 360 VR video model which is similar to those  in  our previous works \cite{guo_tdma,guo_ofdma,long,long2}, and the details are presented here for completeness.} The server is equipped with $M$ transmit antennas and each user wears a single-antenna VR headset. Denote $\mathcal{K}\triangleq\{1,\ldots,K\}$ as the set of user indices.
When a VR user is interested in one viewing direction of a 360 VR video, the user watches a rectangular FoV of size $F_h\times F_v$ (in rad$\times$rad), the center of which is referred to as the viewing direction.
Besides, a user can freely switch views when watching a 360 VR video.

Tiling is adopted to improve the transmission efficiency of the 360 VR video \cite{tiling1}. In particular, the 360 VR video is divided into $U_h\times U_v$ rectangular segments of the same size, referred to as tiles, where $U_h$ and $U_v$ represent the numbers of segments in each row and column, respectively. Define $\mathcal{U}_h\triangleq\{1,\ldots,U_h\}$ and $\mathcal{U}_v\triangleq\{1,\ldots,U_v\}$.
The $(u_h,u_v)$-th tile corresponds to the tile in the $u_h$-th row and the $u_v$-th column, for all $u_h\in\mathcal{U}_h$ and $u_v\in\mathcal{U}_v$.
Considering user heterogeneity (e.g.,  channel conditions, display resolutions of devices,  etc.), each tile is pre-encoded into $L$ representations corresponding to $L$ quality levels using High Efficiency Video Coding (HEVC), as in  Dynamic Adaptive Streaming over HTTP (DASH).
Denote $\mathcal{L}\triangleq\{1,\ldots,L\}$ as the set of quality levels.
For all $l\in\mathcal{L}$, the $l$-th representation of each tile corresponds to the $l$-th lowest quality. For ease of exposition, we assume that the encoding rates of the tiles with the same
quality level are identical.  Let $D_l$ (in bits/s) denote the  encoding rate of the $l$-th representation of a tile.
Note that $D_1<D_2<\ldots<D_L$.
We study the system for the duration of the playback time of multiple groups of pictures (GOPs).\footnote{The duration of the playback time of one GOP is usually $0.06$-$1$ seconds.} In this duration, the FoV of each user does not change. Denote $r_k\in\mathcal{L}$ as the quality level for the FoV of user $k\in\mathcal{K}$.
Because of the video coding structure, $\mathbf{r}\triangleq(r_k)_{k\in\mathcal{K}}$ should not change during the considered time duration.


As in \cite{guo_tdma,guo_ofdma,long,long2}, the server collects a user's information such as head orientation (tracked by a 3DoF or 6DoF VR headset) and location (tracked by a 6DoF VR headset) from the user's headset via the uplink transmission, predicts the user's FoV and determines the set of tiles to be transmitted to the user.\footnote{A widely adopted mechanism for dealing with possible prediction errors is to transmit the tiles in the predicted FoV plus a safe margin \cite{long,long2}. A more significant prediction error yields a larger safe margin, leading to more transmission resource consumption. Note that the proposed framework does not rely on any particular prediction method or transmission mechanism and only focuses on the optimal design of transmitting the scheduled tiles.} Denote $\mathcal{G}_k$ as the set of indices of the tiles that need to be transmitted to user $k$. Denote $\mathcal{G}\triangleq\bigcup_{k\in\mathcal{K}}\mathcal{G}_k$ as the set of indices of the tiles that need to be sent to all $K$ users.
 For all $\mathcal{S}\subseteq\mathcal{K}$, $\mathcal{S}\neq\emptyset$,
 denote $\mathcal{P}_{\mathcal{S}}\triangleq\left(\bigcap_{k\in\mathcal{S}}\mathcal{G}_k
 \right)\bigcap\left(\mathcal{G}-\bigcup_{k\in\mathcal{K}\setminus\mathcal{S}}\mathcal{G}_k\right)$
as the set of indices of the tiles that are needed by all users in $\mathcal{S}$ and are not needed by the users in $\mathcal{K}\setminus\mathcal{S}$.
Then $\mathcal{P}\triangleq\{\mathcal{P}_{\mathcal{S}}|\mathcal{P}_{\mathcal{S}}
\neq\emptyset,~\mathcal{S}\subseteq\mathcal{K},~\mathcal{S}\neq\emptyset\}$ forms
a partition of $\mathcal{G}$
and $\mathcal{I}\triangleq\{\mathcal{S}|\mathcal{P}_{\mathcal{S}}\neq\emptyset,
~\mathcal{S}\subseteq\mathcal{K},~\mathcal{S}\neq\emptyset\}$
specifies the user sets corresponding to the partition. Denote $I \triangleq |\mathcal{I}|$.
Let $\mathcal{I}_k\triangleq\{\mathcal{S}|\mathcal{S}\subseteq\mathcal{I},k\in\mathcal{S}\}$,  $k\in\mathcal{K}$.
The tiles in $\mathcal{P}_{\mathcal{S}}$, $\mathcal{S}\in\mathcal{I}_k$ are required by user $k$.
We consider the tiles in each set jointly  rather than  separately, to
reduce the complexity for transmission and resource allocation.
For all $l\in\mathcal{L}$ and $\mathcal{S}\subseteq\mathcal{K}$, the encoding (source coding) bits of
the $l$-th representations of the tiles in $\mathcal{P}_{\mathcal{S}}$ are ``aggregated'' into one message indexed by $(\mathcal{S},l)$, which is transmitted at most once to the users in $\mathcal{S}$ that will utilize it, to improve transmission efficiency.
For all $\mathcal{S}\in\mathcal{I}$ and $l\in\mathcal{L}$,
let $\mathcal{K}_{\mathcal{S},l}\triangleq\{k\in\mathcal{S}|r_k=l\}$.
If there is only one user in $\mathcal{K}_{\mathcal{S},l}$, the transmission of message $(\mathcal{S},l)$ corresponds to unicast;
and if there are multiple users in $\mathcal{K}_{\mathcal{S},l}$, the transmission of message $(\mathcal{S},l)$ corresponds to multicast.
Thus,  the multi-quality tiled 360 VR video transmission to the $K$ users may involve
both unicast and multicast. An illustration example can be seen in Fig.~\ref{fig:system-model} (b).

Let $\mathcal{N}\triangleq\{1,\ldots,N\}$, where $N$ is the number of subcarriers.
The bandwidth of each subcarrier is $B$ (in Hz).
We assume block fading, i.e., the small-scale channel fading coefficients do not change within one frame.
Let $\mathbf{H}_{n,k}\in\mathbb{C}^{M\times1}$ denote the random
small-scale fading coefficient between the server and user $k$ on subcarrier $n$.
Denote $\mathbf{H}\triangleq (\mathbf{H}_{n,k})_{n\in\mathcal{N},k\in\mathcal K} $.
Let $\mathbf{h}\triangleq (\mathbf{h}_{n,k})_{n\in\mathcal{N},k\in\mathcal K} $ represent a realization of $\mathbf{H}$ (which can be obtained by the server via channel estimation), where $\mathbf{h}_{n,k}\in\mathbb{C}^{M\times1}$ is a realization of $\mathbf{H}_{n,k}$.
Let $\beta_k>0$ denote the large-scale channel fading gain between the server and user $k$, which remains constant during the considered time duration  and is known to the server.

Denote $\mu_{\mathcal{S},l,n}(\mathbf{h})\in\{0,1\}$  as the subcarrier assignment indicator
for subcarrier $n$ and message $(\mathcal{S},l)$ under
$\mathbf{h}$,
where $\mu_{\mathcal{S},l,n}(\mathbf{h})=1$ indicates that subcarrier $n$ is assigned to transmit the symbols for message $(\mathcal{S},l)$,  $\mu_{\mathcal{S},l,n}(\mathbf{h})=0$ otherwise.
For ease of implementation, assume that each subcarrier is assigned to transmit symbols of only one message in each frame \cite{luo2006,choi2015minimum}.
 Thus, subcarrier allocation constraints are given by
\begin{align}
&\mu_{\mathcal{S},l,n}(\mathbf{h}) \in\{0,1\},~\mathcal{S}\in\mathcal{I},l\in\mathcal{L},n\in\mathcal{N},\label{cst:mu1}\\
&\sum\nolimits_{\mathcal{S}\in\mathcal{I}}\sum\nolimits_{l\in\mathcal{L}}\mu_{\mathcal{S},l,n}(\mathbf{h}) =1,~n\in\mathcal{N}.\label{cst:mu2}
\end{align}

To capture the scaling of the transmission power with $M$ for studying the optimal power allocation at large $M$, let $\frac{\eta_{\mathcal{S},l,n}(\mathbf{h})}{M}$ denote the transmission power for  the symbols for message $(\mathcal{S},l)$ on  subcarrier $n$
under $\mathbf{h}$, where
\begin{equation}
\eta_{\mathcal{S},l,n}(\mathbf{h}) \geq0,~\mathcal{S}\in\mathcal{I},l\in\mathcal{L},n\in\mathcal{N}.\label{cst:eta}
\end{equation}
The total transmission power is $\sum\limits_{n\in\mathcal{N}}\sum\limits_{\mathcal{S}\in\mathcal{I}}\sum\limits_{l\in\mathcal{L}}\frac{\mu_{\mathcal{S},l,n}(\mathbf{h}) \eta_{\mathcal{S},l,n}(\mathbf{h})}{M}.$

Suppose that subcarrier $n$ is assigned to transmit the symbols for message $(\mathcal{S},l)$. Let $s_{\mathcal{S},l,n}$ represent the symbols for message $(\mathcal{S},l)$ transmitted on subcarrier $n$.
Assume $\mathbb{E}[|s_{\mathcal{S},l,n}|^2]=1$ for all $l\in\mathcal{L},\mathcal{S}\in\mathcal{I}$ and $n\in\mathcal{N}$.
Let $\mathbf{w}_{n}(\mathbf{h})\in\mathbb{C}^{M\times1}$ denote the  beamforming vector for the message transmitted on subcarrier $n$ under $\mathbf{h}$,
where
\begin{equation}
\|\mathbf{w}_{n}(\mathbf{h}) \|_2=1,~n\in\mathcal{N}.\label{cst:w}
\end{equation}
The received signal at user $k$ on subcarrier $n$ is given by
\begin{equation}
y_{\mathcal{S},l,k,n}=\sqrt{\beta_{k}\frac{\eta_{\mathcal{S},l,n}(\mathbf{h})}{M}}\mathbf{h}_{n,k} ^{H}\mathbf{w}_{n}(\mathbf{h})s_{\mathcal{S},l,n}+z_{n,k},~k\in\mathcal{K},n\in\mathcal{N},\nonumber
\end{equation}

\noindent \normalsize{where $z_{n,k}\sim\mathcal{CN}(0,\sigma^2)$ represents the noise at user $k$ on subcarrier $n$.
To obtain design insights, we consider the application of a capacity achieving code \cite{luo2006,cap2,cap3}. The
maximum transmission rate for the symbols for message $(\mathcal{S},l)$ to user $k\in\mathcal{S}$ on subcarrier $n$ under $\mathbf{h}$ is given by $B\log_2\left(1+
    \frac{\eta_{\mathcal{S},l,n}(\mathbf{h})\beta_{k}|\mathbf{h}_{n,k}^H\mathbf{w}_{n}(\mathbf{h})|^2}
    {M\sigma^2}\right)$ (in bit/s).
Let $c_{\mathcal{S},l,n}(\mathbf{h})$   denote the transmission rate for  the symbols for message $(\mathcal{S},l)$  on  subcarrier $n$ under $\mathbf{h}$,  where}
\begin{equation}
c_{\mathcal{S},l,n}(\mathbf{h})\geq0,~\mathcal{S}\in\mathcal{I},l\in\mathcal{L},n\in\mathcal{N}.\label{cst:c}
\end{equation}
Then, $\sum_{n\in\mathcal{N}}c_{\mathcal{S},l,n}$ represents the transmission rate of message $(\mathcal{S},l)$.

In Section III, we will consider the scenario where users cannot perform transcoding but
 directly play the received messages. In Section IV, we will consider  the scenario where users
can first perform transcoding, i.e., convert a tile representation  at a certain
quality level to  a lower quality level using
transcoding tools such as
Fast Forward Mpeg (FFmpeg), and then play the received video.

\section{Total Transmission Power Minimization Without User Transcoding}\label{section:minp}
In this section, we consider the scenario without user transcoding. For all $\mathcal{S}\in\mathcal{I}$, let $\mathcal{L}_{\mathcal{S}}\triangleq\{r_k|k\in\mathcal{S}\}$. When $|\mathcal{K}_{\mathcal{S},l}| >1$,
message $(\mathcal{S},l)$, where $\mathcal{S}\in\mathcal{I}$ and
$l\in\mathcal{L}_{\mathcal{S}}$
can be transmitted to all users in $\mathcal{K}_{\mathcal{S},l}$ simultaneously via multicast.
This type of multicast opportunities are referred to as natural multicast opportunities \cite{guo_tdma,guo_ofdma,long,long2}.
An  illustration example is shown in Fig.~\ref{fig:system-model} (b). In this example, by using natural multicast opportunities, the server multicasts message $(\{1,2\},1)$ to user 1 and user 2.

Consider one frame with small-scale fading coefficients $\mathbf{h}$.
To guarantee that message $(\mathcal{S},l)$
can be successfully sent to each user $k\in\mathcal{K}_{\mathcal{S},l}$
on subcarrier $n$ under $\mathbf{h}$,  we have
\begin{align}
&\mu_{\mathcal{S},l,n}(\mathbf{h})B\log_2\left(1+
    \frac{\eta_{\mathcal{S},l,n}(\mathbf{h})\beta_{k}|\mathbf{h}_{n,k}^H\mathbf{w}_{n}(\mathbf{h})|^2}
    {M\sigma^2}\right)\nonumber\\
&\geq c_{\mathcal{S},l,n}(\mathbf{h}) ,\mathcal{S}\in\mathcal{I},l\in\mathcal{L}_{\mathcal{S}},k\in\mathcal{K}_{\mathcal{S},l},n\in\mathcal{N}.\label{cst:minrate}
\end{align}
\noindent \normalsize{To maximally avoid rebuffering and reduce startup delay, we require that the transmission rate of each message $(\mathcal{S},l)$ is no smaller than its encoding rate}
\begin{equation}
\sum\nolimits_{n\in\mathcal{N}}c_{\mathcal{S},l,n}(\mathbf{h})\geq |\mathcal{P}_{\mathcal{S}}|D_{l},~\mathcal{S}\in\mathcal{I},l\in
    \mathcal{L}_{\mathcal{S}},\label{cst:averate}
\end{equation}
where $|\mathcal{P}_{\mathcal{S}}|$ denotes the number of tiles in $\mathcal{P}_{\mathcal{S}}$.




For convenience, denote
$\bm{\mu}(\mathbf{h}) \triangleq(\mu_{\mathcal{S},l,n}(\mathbf{h}) )_{\mathcal{S}\in\mathcal{I},l\in\mathcal{L}_{\mathcal{S}},n\in\mathcal{N}}$,
$\bm{\eta}(\mathbf{h}) \triangleq(\eta_{\mathcal{S},l,n}(\mathbf{h}) )_{\mathcal{S}\in\mathcal{I},l\in\mathcal{L}_{\mathcal{S}},n\in\mathcal{N}}$,
 $\mathbf{c}(\mathbf{h}) \triangleq(c_{\mathcal{S},l,n}(\mathbf{h}) )_{\mathcal{S}\in\mathcal{I},l\in\mathcal{L}_{\mathcal{S}},n\in\mathcal{N}}$
 and $\mathbf{w}(\mathbf{h}) \triangleq(\mathbf{w}_{n}(\mathbf{h}) )_{n\in\mathcal{N}}$.
We consider $\bm{\mu}(\mathbf{h}),\bm{\eta}(\mathbf{h}),\mathbf{c}(\mathbf{h}),\mathbf{w}(\mathbf{h})$ as functions of $\mathbf{h}$, respectively.
Given the quality levels of all users $\mathbf{r}$, we would
like to optimize $\bm{\mu}(\mathbf{h})$, $\bm{\eta}(\mathbf{h})$, $\mathbf{c}(\mathbf{h})$, and  $\mathbf{w}(\mathbf{h})$ to  minimize the average total transmission power
\begin{equation}
\mathbb{E}\left[\sum\limits_{n\in\mathcal{N}}\sum\limits_{\mathcal{S}\in\mathcal{I}}\sum\limits_{l\in\mathcal{L}_{\mathcal{S}}}
\frac{\mu_{\mathcal{S},l,n}(\mathbf{H}) \eta_{\mathcal{S},l,n}(\mathbf{H})}{M}\right],\nonumber
\end{equation}
where the expectation is taken over $\mathbf{H}$, subject to the constraints in
\eqref{cst:mu1}-\eqref{cst:w} and \eqref{cst:c}-\eqref{cst:averate}. This is a variational problem due to  the calculus of variation in the objective function. Note that for each $\mathbf{h}$, the number of optimization variables is the same. Also, the optimization variables and constraints are separated for all $\mathbf{h}$. Consequently, it is equivalent to optimize $\bm{\mu}(\mathbf{h})$, $\bm{\eta}(\mathbf{h})$,   $\mathbf{c}(\mathbf{h})$, and  $\mathbf{w}(\mathbf{h})$ to minimize $\frac{1}{M}\sum\nolimits_{n\in\mathcal{N}}\sum\nolimits_{\mathcal{S}\in\mathcal{I}}\sum\nolimits_{l\in\mathcal{L}_{\mathcal{S}}}\mu_{\mathcal{S},l,n}(\mathbf{h}) \eta_{\mathcal{S},l,n}(\mathbf{h})$ subject to the constraints in
\eqref{cst:mu1}-\eqref{cst:w}, and \eqref{cst:c}-\eqref{cst:averate},  for all $\mathbf{h}$. Thus, we consider the following problem.
\begin{Prob}[Total Transmission Power Minimization  for $\mathbf{h} $]\label{P1}
\begin{align}
&E^{\star}(\mathbf{h})\triangleq\min_{\bm{\mu}(\mathbf{h}),\bm{\eta}(\mathbf{h}),\mathbf{c}(\mathbf{h}),\mathbf{w}(\mathbf{h})}~
\sum\limits_{n\in\mathcal{N}}\sum\limits_{\mathcal{S}\in\mathcal{I}}\sum\limits_{l\in\mathcal{L}_{\mathcal{S}}}
\frac{\mu_{\mathcal{S},l,n}(\mathbf{h}) \eta_{\mathcal{S},l,n}(\mathbf{h})}{M}\nonumber\\
    &\quad\quad\quad\quad\mathrm{s.t.} ~~
\eqref{cst:mu1},~\eqref{cst:mu2},~\eqref{cst:eta},~\eqref{cst:w},~\eqref{cst:c},~\eqref{cst:minrate},~\eqref{cst:averate}.\nonumber
\end{align}
\end{Prob}

By noting that $\bm{\mu}(\mathbf{h})$,
$\bm{\eta}(\mathbf{h})$, and
$\mathbf{c}(\mathbf{h})$ represent the subcarrier allocation, transmission power allocation, and transmission rate of all messages $(\mathcal{S},l),\mathcal{S}\in\mathcal{I},l\in\mathcal{L}_{\mathcal{S}}$ on all subcarriers $n\in\mathcal{N}$, the overlapping of the FoVs of the users and the quality requirements of the users are captured in Problem~1. Therefore, Problem~1 reflects the wireless streaming of the multi-quality tiled 360 VR video to the $K$ users.
It can be observed that Problem~\ref{P1} is a challenging mixed discrete-continuous optimization problem.
In Section~\ref{sec:p1-sp}, we first obtain a globally optimal solution for small multicast groups and an asymptotically optimal solution for a large antenna array. Then, in Section~\ref{sec:p1-dc}, we obtain a suboptimal solution for the general case.\footnote{Note that the goal of solving a nonconvex problem is usually to design an iterative algorithm to obtain a stationary point or a KKT point. In general, it is impossible to analytically or numerically show the gap between a globally optimal solution and a suboptimal solution, as a globally optimal solution cannot be obtained analytically or numerically with effective and efficient methods \cite{NL}.}


\subsection{Solutions in Special Cases}\label{sec:p1-sp}
In this subsection, we
solve Problem~\ref{P1} in two special cases, by solving the following equivalent problem of Problem~\ref{P1}.

%

\begin{Prob}[Equivalent Problem of Problem~\ref{P1} for $\mathbf{h}$]\label{EP1}
\begin{align}
&E^{\dagger}(\mathbf{h})\triangleq\min_{\bm{\mu}(\mathbf{h}),\mathbf{P}(\mathbf{h}) }~\frac{1}{M}\sum\limits_{n\in\mathcal{N}}\sum\limits_{\mathcal{S}\in\mathcal{I}}\sum\limits_{l\in\mathcal{L}_{\mathcal{S}}}P_{\mathcal{S},l,n}(\mathbf{h}) \nonumber\\
    &\mathrm{s.t.} ~~
    \eqref{cst:mu1},~\eqref{cst:mu2},\nonumber\\
    &P_{\mathcal{S},l,n}(\mathbf{h})\geq0,~\mathcal{S}\in\mathcal{I},l\in\mathcal{L}_{\mathcal{S}},n\in\mathcal{N},\label{cst:p}\\
&\sum\limits_{n\in\mathcal{N}}\mu_{\mathcal{S},l,n}(\mathbf{h})B\log_2\left(1+
    \frac{P_{\mathcal{S},l,n}(\mathbf{h})}
    {\mu_{\mathcal{S},l,n}(\mathbf{h})Q_{\mathcal{S},l,n}^{\dagger}(\mathbf{h})}\right)\nonumber\\
    &\geq |\mathcal{P}_{\mathcal{S}}|D_{l},~\mathcal{S}\in\mathcal{I},l\in\mathcal{L}_{\mathcal{S}},k\in\mathcal{K}_{\mathcal{S},l},\label{cst:sumraten}
\end{align}
where $\mathbf{P}(\mathbf{h})\triangleq(P_{\mathcal{S},l,n}(\mathbf{h}))_{\mathcal{S}\in\mathcal{I},l\in\mathcal{L}_{\mathcal{S}},n\in\mathcal{N}}$
and $Q_{\mathcal{S},l,n}^{\dagger}(\mathbf{h})$ is given by the following problem.
Let $(\bm{\mu}^{\dagger}(\mathbf{h}) ,\mathbf{P}^{\dagger}(\mathbf{h}) )$ denote an optimal solution of Problem~\ref{EP1}.
\end{Prob}
\begin{Prob}[Subproblem of Problem~\ref{EP1} for $\mathbf{h}$]\label{BF1}
\begin{align}
Q_{\mathcal{S},l,n}^{\dagger}(\mathbf{h})\triangleq&\min_{\mathbf{V}_{\mathcal{S},l,n}\in\mathbb{C}^{M\times M}}\text{tr}(\mathbf{V}_{\mathcal{S},l,n})
    \nonumber\\
    \mathrm{s.t.} ~~&\frac{\text{tr}(\beta_k\mathbf{h}_{k,n} \mathbf{h}_{k,n} ^H\mathbf{V}_{\mathcal{S},l,n})}{M\sigma^2}\geq1,~k\in\mathcal{K}_{\mathcal{S},l},\label{cst:bf1}\\
    &\mathbf{V}_{\mathcal{S},l,n}\succeq\mathbf{0},\nonumber\\
    &\text{rank}(\mathbf{V}_{\mathcal{S},l,n})=1.\label{cst:rank1}
\end{align}
Let $\mathbf{V}_{\mathcal{S},l,n}^{\dagger}(\mathbf{h}) $ denote an optimal solution of Problem~\ref{BF1},
which can be written as $\mathbf{V}_{\mathcal{S},l,n}^{\dagger}(\mathbf{h}) =\mathbf{v}^{\dagger}_{\mathcal{S},l,n}(\mathbf{h})
(\mathbf{v}^{\dagger}_{\mathcal{S},l,n}(\mathbf{h}) )^H$ for some $\mathbf{v}^{\dagger}_{\mathcal{S},l,n}(\mathbf{h}) \in\mathbb{C}^{M\times1}$.
\end{Prob}

By exploring structures of Problem~\ref{P1}, Problem~\ref{EP1}, and Problem~\ref{BF1}, we have the following result.
\begin{Thm}[Equivalence between Problem~\ref{P1} and Problem~\ref{EP1}]\label{lem:eqp1}
The optimal values of Problem~\ref{P1} and Problem~\ref{EP1} are identical.
In addition, $(\bm{\mu}^{\dagger}(\mathbf{h}),\bm{\eta}^{\dagger}(\mathbf{h}),\mathbf{c}^{\dagger}(\mathbf{h}),
\mathbf{w}^{\dagger}(\mathbf{h}))$
is an optimal solution of Problem~\ref{P1}, where
\begin{align}
&\bm{\eta}^{\dagger}(\mathbf{h})=\mathbf{P}^{\dagger}(\mathbf{h}),\label{thm1-eta}\\
&\mathbf{w}^{\dagger}_{n}(\mathbf{h}) =\sum\limits_{\mathcal{S}\in\mathcal{I}}\sum\limits_{l\in\mathcal{L}_{\mathcal{S}}}\mu_{\mathcal{S},l,n}^{\dagger}(\mathbf{h})\frac{\mathbf{v}^{\dagger}_{\mathcal{S},l,n}(\mathbf{h}) }{\sqrt{Q_{\mathcal{S},l,n}^{\dagger}(\mathbf{h})}},~n\in\mathcal{N},\label{thm1-w}
\end{align}
and $\mathbf{c}^{\dagger}(\mathbf{h})\triangleq(c^{\dagger}_{\mathcal{S},l,n}(\mathbf{h}))_{\mathbf{S}\in\mathcal{I},l\in\mathcal{L}_{\mathcal{S}},n\in\mathcal{N}}$
with
\begin{align}
&c_{\mathcal{S},l,n}^{\dagger}(\mathbf{h})=\mu_{\mathcal{S},l,n}^{\dagger}(\mathbf{h})B\log_2\left(1+
    \frac{P_{\mathcal{S},l,n}^{\dagger}(\mathbf{h}) }
    {Q_{\mathcal{S},l,n}^{\dagger}(\mathbf{h}) }\right).\label{thm1-c}
\end{align}
\end{Thm}
\begin{Proof}
 See Appendix~A.
\end{Proof}

According to Theorem~\ref{lem:eqp1}, to obtain an optimal solution of Problem~\ref{P1},
we can first get $\mathbf{w}^{\dagger}(\mathbf{h})$ by solving Problem~\ref{BF1}, and then get $\bm{\mu}^{\dagger}(\mathbf{h})$, $\bm{\eta}^{\dagger}(\mathbf{h})$, and $\mathbf{c}^{\dagger}(\mathbf{h})$ by solving
Problem~\ref{EP1}.
Notice that
 Problem~\ref{BF1} is nonconvex
due to the rank-one constraint in  \eqref{cst:rank1}, while
Problem~\ref{EP1} is a nonconvex problem because of the binary constraints in
\eqref{cst:mu1}.
Both problems are pretty challenging. In the following, we solve Problem~\ref{BF1}
and Problem~\ref{EP1} for two special cases.
\subsubsection{Case of Small Multicast Groups}\label{sec:p1-sp1}
In this part, we consider the case where for all $\mathcal{S}\in{\mathcal{I}}$ and $l\in\mathcal{L}_{\mathcal{S}}$,  there are at most three users who need message $(\mathcal{S},l)$ (i.e.,
$|\mathcal{K}_{\mathcal{S},l}|\leq3$).
First, we obtain an optimal solution of Problem~\ref{BF1}
by applying semidefinite relaxation  and rank reduction proposed in \cite{reduction}.
Specifically, we relax the constraint in \eqref{cst:rank1}, and get an~SDP, which is convex and can be solved effectively. Under the condition that $|\mathcal{K}_{\mathcal{S},l}|\leq3$,
 a rank-one solution can be constructed based on an optimal solution of the SDP using rank reduction \cite{reduction}.
Then, substituting $Q_{\mathcal{S},l,n}^{\dagger}(\mathbf{h})$,
$\mathcal{S}\in\mathcal{I}$, $l\in\mathcal{L}_{\mathcal{S}}$ into Problem~\ref{EP1} and
 relaxing the constraints in \eqref{cst:mu1} to
\begin{equation}
\mu_{\mathcal{S},l,n}(\mathbf{h})\geq 0,~~\mathcal{S}\in\mathcal{I},l\in\mathcal{L}_{\mathcal{S}},n\in\mathcal{N}, \label{const:ofdma-relax}
\end{equation}
we obtain a relaxed problem of Problem~\ref{EP1}, which is convex.
Using the KKT conditions, we know that under a mild condition, there
exists an optimal solution of the relaxed problem of Problem~\ref{EP1} which provides binary subcarrier assignment \cite{guo_ofdma}.
For all $\mathcal{S}\in\mathcal{I},~l\in\mathcal{L}_{\mathcal{S}}$ and $n\in\mathcal{N}$, define:

\begin{align}
&f_{\mathcal{S},l,n}(\mathbf{h},\lambda_{\mathcal{S},l})\triangleq \left[\frac{B\lambda_{\mathcal{S},l}}{\ln2}-Q_{\mathcal{S},l,n}^{\dagger}(\mathbf{h}) \right]^{+},\\
&\mathcal{W}_{\mathcal{S},l,n}(\mathbf{h},\lambda_{\mathcal{S},l})\triangleq \lambda_{\mathcal{S},l}B
\left(\log_2\left(1
+\frac{f_{\mathcal{S},l,n}(\mathbf{h},\lambda_{\mathcal{S},l})}{Q_{\mathcal{S},l,n}^{\dagger}(\mathbf{h})}\right)
\right.\nonumber\\
&\left.-\frac{ f_{\mathcal{S},l,n}(\mathbf{h},\lambda_{\mathcal{S},l})}{\left(Q_{\mathcal{S},l,n}^{\dagger}(\mathbf{h}) + f_{\mathcal{S},l,n}(\mathbf{h},\lambda_{\mathcal{S},l})\right)\ln2}\right),\label{eq:W}\\
&\mu_{\mathcal{S},l,n}(\mathbf{h},\lambda_{\mathcal{S},l})\triangleq
             \begin{cases}
             1, ~ (\mathcal{S},l)=\argmax\limits_{\mathcal{S}'\in\mathcal{I},l'\in\mathcal{L}}\mathcal{W}_{\mathcal{S}',l',n}(\mathbf{h},\lambda_{\mathcal{S}',l'}), \\
             0, ~ \text{otherwise},
             \end{cases}\label{eq:mu}\\
&P_{\mathcal{S},l,n}(\mathbf{h},\lambda_{\mathcal{S},l})\triangleq\mu_{\mathcal{S},l,n}(\mathbf{h},\lambda_{\mathcal{S},l})
f_{\mathcal{S},l,n}(\mathbf{h},\lambda_{\mathcal{S},l}).\label{eq:P}
\end{align}
\normalsize{Let $\lambda^{\dagger}_{\mathcal{S},l}(\mathbf{h})$ denote
a root of}
\begin{align}
&\sum\limits_{n\in\mathcal{N}}\mu_{\mathcal{S},l,n}(\mathbf{h},\lambda_{\mathcal{S},l})B\log_2\left(1+
    \frac{P_{\mathcal{S},l,n}(\mathbf{h},\lambda_{\mathcal{S},l})}
    {\mu_{\mathcal{S},l,n}(\mathbf{h},\lambda_{\mathcal{S},l})Q_{\mathcal{S},l,n}^{\dagger}(\mathbf{h})}\right)
    \nonumber\\
    &=|\mathcal{P}_{\mathcal{S}}|D_{l},~\mathcal{S}\in\mathcal{I},l\in\mathcal{L}_{\mathcal{S}}.
\end{align}

\noindent\normalsize{An optimal solution of Problem~\ref{EP1} is given below \cite{guo_ofdma}.}
\begin{Cla}[Optimal Solution of Problem~\ref{EP1} for $\mathbf{h}$]\label{lem:opt-ofdma}
Suppose that for all $n\in\mathcal{N}$, there exists a unique pair $(\mathcal{S}_n,l_n)$ such that $\mathcal{W}_{\mathcal{S}_n,l_n,n}(\mathbf{h},\lambda^{\dagger}_{\mathcal{S}_n,l_n}(\mathbf{h}) )=\max_{\mathcal{S}\in\mathcal{I},l\in\mathcal{L}_{\mathcal{S}}}\mathcal{W}_{\mathcal{S},l,n}(\mathbf{h},\lambda^{\dagger}_{\mathcal{S},l}(\mathbf{h}) )$. Then,  an optimal solution of Problem~\ref{EP1} is given by
    $\mu_{\mathcal{S},l,n}^{\dagger}(\mathbf{h}) =\mu_{\mathcal{S},l,n}(\mathbf{h},\lambda_{\mathcal{S},l}^{\dagger}(\mathbf{h}))$,
    $P_{\mathcal{S},l,n}^{\dagger}(\mathbf{h}) =P_{\mathcal{S},l,n}(\mathbf{h},\lambda_{\mathcal{S},l}^{\dagger}(\mathbf{h}))$,  $\mathcal{S}\in\mathcal{I}$, $l\in\mathcal{L}_{\mathcal{S}}$ and $n\in\mathcal{N}$.
\end{Cla}

Note that $\mathcal{W}_{\mathcal{S},l,n}(\mathbf{h},\lambda^{\dagger}_{\mathcal{S},l}(\mathbf{h}))$ monotonically
increases with $Q_{\mathcal{S},l,n}^{\dagger}(\mathbf{h})$ and $Q_{\mathcal{S},l,n}^{\dagger}(\mathbf{h})$,
$\mathcal{S}\in\mathcal{I}$, $l\in\mathcal{L}_{\mathcal{S}}$
are different for user groups  $\mathcal{K}_{\mathcal{S},l}$, $\mathcal{S}\in\mathcal{I}$, $l\in\mathcal{L}_{\mathcal{S}}$ (as $Q_{\mathcal{S},l,n}^{\dagger}(\mathbf{h})$ captures both large-scale fading and small-scale fading effects). Thus, $\mathcal{W}_{\mathcal{S},l,n}(\mathbf{h},\lambda_{\mathcal{S},l}^{\dagger}(\mathbf{h}))$, $\mathcal{S}\in\mathcal{I}$, $l\in\mathcal{L}_{\mathcal{S}}$ are usually different, and the condition in Claim~1 can be easily satisfied \cite{guo_ofdma}.
Note that $\lambda^{\dagger}_{\mathcal{S},l}(\mathbf{h})$ can be obtained using a subgradient method as in \cite{long}.

The details for obtaining an optimal solution of Problem~\ref{P1} via solving Problem~\ref{BF1} and Problem~\ref{EP1} are summarized in Algorithm~\ref{alg:opt-GP1}.
Specifically, in Steps 1-13, we solve Problem~\ref{BF1} for all $\mathcal{S}\in\mathcal{I},l\in\mathcal{L}_{\mathcal{S}}$, and $n\in\mathcal{N}$ with computational complexity $\mathcal{O}(M^{6}NKI)$;
in Steps 14-20, we solve Problem~\ref{EP1} with computational complexity  $\mathcal{O}(NLKI^{2})$; in Steps 21-25, we compute the optimal solution of Problem 1 based on the solutions of Problem 2 and Problem 3 with computational complexity  $\mathcal{O}(NKI)$. Therefore, the computational complexity of Algorithm 1 is $\mathcal{O}(NLKI^{2})$.



\small{\begin{algorithm}
    \caption{\small{Globally Optimal Solution of Problem~\ref{P1} for Case of Small Multicast Groups}}
\begin{footnotesize}
     \begin{algorithmic}[1]
     \FOR{$\mathcal{S}\in\mathcal{I},l\in\mathcal{L}_{\mathcal{S}}$ and $n\in\mathcal{N}$}
                      \STATE   Find an optimal solution $\mathbf{V}^{\dagger}_{\mathcal{S},l,n}(\mathbf{h})$ (with arbitrary ranks) of Problem~\ref{BF1} without the rank-one constraint in \eqref{cst:rank1};
        \WHILE{$\text{rank}(\mathbf{V}^{\dagger}_{\mathcal{S},l,n}(\mathbf{h}))>1$}
      \STATE Set $\psi=\text{rank}(\mathbf{V}^{\dagger}_{\mathcal{S},l,n}(\mathbf{h}))$;
      \STATE    Decompose $\mathbf{V}^{\dagger}_{\mathcal{S},l,n}(\mathbf{h})=\mathbf{U}\mathbf{U}^H$;\\
      \STATE    Find a nonzero solution $\mathbf{\Delta}$ of the system of linear equations:
      $\text{tr}(\mathbf{U}^{H}\beta_k\mathbf{h}_{k,n}\mathbf{h}_{k,n}^H\mathbf{U}\mathbf{\Delta})=0$, $k\in\mathcal{K}_{\mathcal{S},l}$,
      where $\mathbf{\mathbf{\Delta}}$ is a $\psi\times\psi$ Hermitian matrix;
      \STATE    Evaluate the eigenvalues $\delta_{1},\ldots,\delta_{\psi}$ of $\mathbf{\Delta}$;\\
      \STATE    Determine $i_{0}$ such that
      $|\delta_{i_{0}}|=\text{max}\{|\delta_{i}|:1\leq i\leq\psi\}$;\\
      \STATE    Compute $\mathbf{V}^{\dagger}_{\mathcal{S},l,n}(\mathbf{h})=\mathbf{U}(\mathbf{I}_{\psi}-(1/\delta_{i_{0}})\mathbf{\Delta})\mathbf{U}^H$;\\
      \ENDWHILE
      \STATE Compute $Q_{\mathcal{S},l,n}^{\dagger}(\mathbf{h})=\text{tr}(\mathbf{V}^{\dagger}_{\mathcal{S},l,n}(\mathbf{h}))$;
      \STATE Decompose $\mathbf{V}^{\dagger}_{\mathcal{S},l,n}(\mathbf{h})=\mathbf{v}_{\mathcal{S},l,n}^{\dagger}(\mathbf{h})
      (\mathbf{v}_{\mathcal{S},l,n}^{\dagger}(\mathbf{h}))^{H}$;
      \ENDFOR
           \STATE Initialize $\bm{\lambda}^{(0)}$. Set iteration index $t=0$;
           \REPEAT
           \STATE  For all $\mathcal{S}\in\mathcal{I},l\in\mathcal{L}_{\mathcal{S}}$ and $n\in\mathcal{N}$, compute $W_{\mathcal{S},l,n}(\mathbf{h},\lambda_{\mathcal{S},l}^{(t)})$ according to \eqref{eq:W};
           \STATE For all $\mathcal{S}\in\mathcal{I},l\in\mathcal{L}_{\mathcal{S}}$ and $n\in\mathcal{N}$, compute $\mu_{\mathcal{S},l,n}(\mathbf{h},\lambda^{(t)}_{\mathcal{S},l})$ and $P_{\mathcal{S},l,n}(\mathbf{h},\lambda^{(t)}_{\mathcal{S},l})$ according to \eqref{eq:mu} and
           \eqref{eq:P}, respectively;
           \STATE For all $\mathcal{S}\in\mathcal{I}$ and $l\in\mathcal{L}_{\mathcal{S}}$, compute $\lambda^{(t+1)}_{\mathcal{S},l}$ according to:
\footnotesize{\begin{align*}
\lambda_{\mathcal{S},l}^{(t+1)}=&\left[\lambda_{\mathcal{S},l}^{(t)}-\delta^{(t)}\left(\sum\nolimits_{n\in\mathcal{N}}\mu_{\mathcal{S},l,n}(\mathbf{h},\lambda^{(t)}_{\mathcal{S},l})
B\log_2\left(1+\right.\right.\right.\nonumber\\
&\left.\left.\left.\frac{P_{\mathcal{S},l,n}(\mathbf{h},\lambda^{(t)}_{\mathcal{S},l})}{\mu_{\mathcal{S},l,n}(\mathbf{h},\lambda^{(t)}_{\mathcal{S},l})Q_{\mathcal{S},l,n}^{\dagger}(\mathbf{h})}\right)
-|\mathcal{P}_{\mathcal{S}}|D_{l}\right)\right]^{+},
\end{align*}}
\noindent where $\delta^{(t)}$, $t=1,2,\ldots$
satisfy
 \begin{equation}
 \delta^{(t)}>0,~\sum\limits_{t=0}^{\infty}(\delta^{(t)})^2<\infty,~\sum\limits_{t=0}^{\infty}\delta^{(t)}=\infty,~\lim\limits_{t\rightarrow\infty}\delta^{(t)}=0;\label{cst:step3}
 \end{equation}
\STATE Set $t=t+1$;
           \UNTIL{convergence criteria is met}
           \FOR{ $n\in\mathcal{N}$}
      		\STATE Compute $\mathbf{w}_{n}^{\dagger}(\mathbf{h})$ according to \eqref{thm1-w};
      		\ENDFOR
           \STATE For all $\mathcal{S}\in\mathcal{I},l\in\mathcal{L}_{\mathcal{S}}$ and $n\in\mathcal{N}$, set $\mu_{\mathcal{S},l,n}^{\dagger}(\mathbf{h})=\mu_{\mathcal{S},l,n}(\mathbf{h},\lambda^{(t)}_{\mathcal{S},l})$
           and $P_{\mathcal{S},l,n}^{\dagger}(\mathbf{h})=P_{\mathcal{S},l,n}(\mathbf{h},\lambda^{(t)}_{\mathcal{S},l})$;
           \STATE
           Set $\bm{\eta}^{\dagger}(\mathbf{h})=\mathbf{P}^{\dagger}(\mathbf{h})$, and compute $\mathbf{c}^{\dagger}(\mathbf{h})$ according to (13).
    \end{algorithmic}
    \end{footnotesize}\label{alg:opt-GP1}
\end{algorithm}}


\subsubsection{Case of a Large Antenna Array}\label{sect:p1-sp2}
\normalsize{In} this part, we consider the case where the server is equipped with  a
large antenna array. For the sake of presentation, in this part,
we explicitly write the optimal value of Problem~\ref{BF1}
as a function of $M$, i.e., $Q_{\mathcal{S},l,n}^{\dagger(M)}(\mathbf{h}) $.
Following the proofs for Theorem~1 and Theorem~3  in \cite{xiang}, we can show the following result.

\begin{Thm}[Asymptotically Optimal Solution of Problem~\ref{BF1}]\label{thm:M}
For all $\mathcal{S}\in\mathcal{I}$, $l\in\mathcal{L}_{\mathcal{S}}$,
$n\in\mathcal{N}$ and $\mathbf{h} $, $\mathbf{V}_{\mathcal{S},l,n}^{\ast}(\mathbf{h})=\mathbf{v}_{\mathcal{S},l,n}^{\ast}(\mathbf{h})(\mathbf{v}_{\mathcal{S},l,n}^{\ast}(\mathbf{h}))^H$ is
    an asymptotically optimal solution of Problem~\ref{BF1} at large $M$, where
\begin{align}
&\mathbf{v}_{\mathcal{S},l,n}^{\ast}(\mathbf{h})=\frac{\sum\limits_{k\in\mathcal{K}_{\mathcal{S},l}}\frac{\mathbf{h}_{n,k}}{\sqrt{\beta_k}}}{\left\|\sum\limits_{k\in\mathcal{K}_{\mathcal{S},l}}\frac{\mathbf{h}_{n,k}}{\sqrt{\beta_k}}\right\|_2} \sqrt{\frac{M\sigma^2}{\min\limits_{k\in\mathcal{K}_{\mathcal{S},l}}
\beta_k\frac{\left|\sum\limits_{j\in\mathcal{K}_{\mathcal{S},l}}\frac{\mathbf{h}_{n,k}^H\mathbf{h}_{n,j}}{\sqrt{\beta_j}}\right|^2}{\left\|\sum\limits_{j\in\mathcal{K}_{\mathcal{S},l}}\frac{\mathbf{h}_{n,j}}{\sqrt{\beta_j}}\right\|_2^2}}}.
\label{thm2-v}
\end{align}
\end{Thm}

\normalsize{\begin{Proof}
 See Appendix~B.
\end{Proof}}


Substituting $Q_{\mathcal{S},l,n}^{\dagger(M)}(\mathbf{h})=\text{tr}(\mathbf{V}_{\mathcal{S},l,n}^{\ast}(\mathbf{h}))$  into Problem~\ref{EP1}
and using the same method as in Section~\ref{sec:p1-sp1} for solving Problem~\ref{EP1},
 an asymptotically optimal solution of Problem~\ref{P1} (which can achieve competitive performance at large $M$) can be obtained.
\subsection{Suboptimal Solution in General Case}\label{sec:p1-dc}
In the general case, i.e., there exists $(\mathcal{S},l)$ such that $|\mathcal{K}_{\mathcal{S},l}|>3$ and the number of antennas equipped at the server is not large, we cannot obtain a globally optimal solution of the nonconvex problem in Problem~\ref{BF1}. Thus, we cannot solve Problem~\ref{P1} by solving its equivalent form in Problem~\ref{EP1}.
In this subsection, we directly tackle Problem~\ref{P1}, and develop a low-complexity algorithm to obtain a suboptimal solution of Problem~\ref{P1} using relaxation and  DC programming.

First, by relaxing the constraints in \eqref{cst:mu1} of Problem~1 to the  constraints in \eqref{const:ofdma-relax},
 we can obtain the relaxed continuous problem of Problem~\ref{P1}.
Next, we convert the relaxed continuous problem of Problem~\ref{P1} to  DC programming.
Let $\mathbf{W}_{\mathcal{S},l,n}(\mathbf{h})\triangleq\sqrt{\eta_{\mathcal{S},l,n}(\mathbf{h}) \mu_{\mathcal{S},l,n}(\mathbf{h})}\mathbf{w}_{n}(\mathbf{h})$.
Thus, the constraints in \eqref{cst:eta}, \eqref{cst:w}, and \eqref{cst:minrate} can be equivalently transformed to the following ones.
\begin{align}
  &\mu_{\mathcal{S},l,n}(\mathbf{h})\left(2^{\frac{c_{\mathcal{S},l,n}(\mathbf{h}) }{B\mu_{\mathcal{S},l,n}(\mathbf{h}) }}-1\right)-\frac{\beta_k|\mathbf{h}_{n,k} ^H\mathbf{W}_{\mathcal{S},l,n}(\mathbf{h})|^2}{M\sigma^2}\leq 0,\nonumber\\
&\mathcal{S}\in\mathcal{I},l\in\mathcal{L}_{\mathcal{S}},k\in\mathcal{K}_{\mathcal{S},l},n\in\mathcal{N}.\label{cst:minraten}
\end{align}
Thus, the relaxed continuous problem of Problem~\ref{P1} is given as follows.
\begin{Prob}[DC Problem of Relaxed Problem~\ref{P1} for $\mathbf{h}$]\label{RP1}
\begin{align}
 \min_{\mathbf{W}(\mathbf{h}),\bm{\mu}(\mathbf{h}),\mathbf{c}(\mathbf{h}) }~&\frac{1}{M}\sum\limits_{n\in\mathcal{N}}\sum\limits_{\mathcal{S}\in\mathcal{I}}\sum\limits_{l\in\mathcal{L}_{\mathcal{S}}}\|\mathbf{W}_{\mathcal{S},l,n}(\mathbf{h}) \|_2^2
 \nonumber\\
    \mathrm{s.t.} ~~
    &\eqref{cst:mu2},~\eqref{cst:c},~\eqref{cst:averate},~\eqref{const:ofdma-relax},~\eqref{cst:minraten}.\nonumber
\end{align}
\end{Prob}

Note that the objective function of Problem~\ref{RP1}
and the constraints in \eqref{cst:mu2},~\eqref{cst:c},~\eqref{cst:averate}, and \eqref{const:ofdma-relax} are all convex.
Besides,
each of the constraints in \eqref{cst:minraten}
can be regarded as a difference of two convex functions,
i.e., $\mu_{\mathcal{S},l,n}(\mathbf{h})\left(2^{\frac{c_{\mathcal{S},l,n}(\mathbf{h}) }{B\mu_{\mathcal{S},l,n}(\mathbf{h}) }}-1\right)$ and $\frac{\beta_k|\mathbf{h}_{n,k} ^H\mathbf{W}_{\mathcal{S},l,n}(\mathbf{h})|^2}{M\sigma^2}$.
Thus, Problem~\ref{RP1} is a standard DC programming and can be handled by using the DC algorithm \cite{dc}. The core idea
is to  solve a sequence of convex approximations of Problem~\ref{RP1} iteratively, each of which is obtained by linearizing
 the concave function,
 i.e., $-\frac{\beta_k|\mathbf{h}_{n,k}^H\mathbf{W}_{\mathcal{S},l,n}(\mathbf{h}) |^2}{M\sigma^2}$ in \eqref{cst:minraten}.
Specifically, at the $t$-th iteration,  the convex approximation of Problem~\ref{RP1}
is given below.
\begin{Prob}[Convex Approximation of Problem~\ref{RP1} for $\mathbf{h}$ \protect\ at
$t$-th Iteration]\label{CRP1}
\begin{align}
 E^{(t)}(\mathbf{h})\triangleq\min_{\mathbf{W}(\mathbf{h}),\bm{\mu}(\mathbf{h}),\mathbf{c}(\mathbf{h}) }~&\frac{1}{M}\sum\limits_{n\in\mathcal{N}}\sum\limits_{\mathcal{S}\in\mathcal{I}}\sum\limits_{l\in\mathcal{L}}\|\mathbf{W}_{\mathcal{S},l,n}(\mathbf{h}) \|^2 \nonumber\\
    \mathrm{s.t.} ~
    \eqref{cst:mu2},~&\eqref{cst:c},~\eqref{cst:averate},~\eqref{const:ofdma-relax},~\eqref{cst:linear},\nonumber
\end{align}
where \eqref{cst:linear} is shown at the top of the next page.
Let $(\mathbf{W}^{(t)}(\mathbf{h}) ,\bm{\mu}^{(t)}(\mathbf{h}) ,\mathbf{c}^{(t)}(\mathbf{h}) )$ denote an optimal solution at the $t$-th iteration.
\end{Prob}

\begin{figure*}[!t]
\small{
\begin{align}
    \mu_{\mathcal{S},l,n}(\mathbf{h})\left(2^{\frac{c_{\mathcal{S},l,n}(\mathbf{h}) }{B\mu_{\mathcal{S},l,n}(\mathbf{h}) }}-1\right)-\frac{2\beta_kR\left\{(\mathbf{W}^{(t-1)}_{\mathcal{S},l,n}(\mathbf{h}) )^H\mathbf{h}_{n,k} \mathbf{h}_{n,k} ^H\mathbf{W}_{\mathcal{S},l,n}(\mathbf{h})\right\}}{M\sigma^2}+\frac{|\mathbf{h}_{n,k}^H\mathbf{W}^{(t-1)}_{\mathcal{S},l,n}(\mathbf{h}) |^2}{M\sigma^2}\leq0,~\mathcal{S}\in\mathcal{I},l\in\mathcal{L}_{\mathcal{S}},k\in\mathcal{K}_{\mathcal{S},l},n\in\mathcal{N}.\label{cst:linear}\end{align}
}
\end{figure*}


Problem~\ref{CRP1} is a convex problem. We can use standard convex optimization techniques to solve it.
According to \cite{dc}, for any initial point which is a feasible solution of Problem~\ref{RP1},
as $t\rightarrow\infty$, $(\mathbf{W}^{(t)}(\mathbf{h}) ,\bm{\mu}^{(t)}(\mathbf{h}) ,\mathbf{c}^{(t)}(\mathbf{h}) )\rightarrow(\mathbf{W}^{(\infty)}(\mathbf{h}) ,\bm{\mu}^{(\infty)}(\mathbf{h}) ,\mathbf{c}^{(\infty)}(\mathbf{h}) )$, which
 is a stationary point of the relaxed Problem~\ref{P1}, and $E^{(t)}(\mathbf{h})\rightarrow E^{(\infty)}(\mathbf{h})$.
Note that $\bm{\mu}^{(\infty)}(\mathbf{h})$ may not be binary, and hence
 $(\mathbf{W}^{(\infty)}(\mathbf{h}) ,\bm{\mu}^{(\infty)}(\mathbf{h}) ,\mathbf{c}^{(\infty)}(\mathbf{h}) )$ may not be a feasible solution of Problem~\ref{P1}.
 By using the KKT conditions, we can obtain an optimal solution of  Problem~\ref{CRP1}
 for the $t^{\diamond}$-the iteration where $t^{\diamond}$ satisfies some
 convergence criteria. It provides binary subcarrier assignment under a mild condition, and can be treated as a suboptimal solution of Problem~\ref{P1}.

 Let $\bm{\lambda}_{\mathcal{S},l,n}\triangleq(\lambda_{\mathcal{S},l,n,k})_{k\in\mathcal{K}_{\mathcal{S},l}}$.
For all $\mathcal{S}\in\mathcal{I}$, $l\in\mathcal{L}_{\mathcal{S}}$ and
$n\in\mathcal{N}$, define:
\small{\begin{align}
&G_{\mathcal{S},l,n}(\mathbf{h},\gamma_{\mathcal{S},l},\bm{\lambda}_{\mathcal{S},l,n})\triangleq
\gamma_{\mathcal{S},l}\log_2\frac{\gamma_{\mathcal{S},l}}
{\ln2\sum\nolimits_{k\in\mathcal{S}}\lambda_{\mathcal{S},l,n,k}}-\frac{\gamma_{\mathcal{S},l}B}{\ln2}\nonumber\\
&+\sum\nolimits_{k\in\mathcal{S}}\lambda_{\mathcal{S},l,n,k},\label{cst:dc-g}\\
&\mu_{\mathcal{S},l,n}(\mathbf{h},\gamma_{\mathcal{S},l},\bm{\lambda}_{\mathcal{S},l,n})
\nonumber\\&=\begin{cases}
             1, ~ (\mathcal{S},l)=\argmax\limits_{\mathcal{S}'\in\mathcal{I},l'\in\mathcal{L}_{\mathcal{S}}}G_{\mathcal{S}',l',n}(\mathbf{h},\gamma_{\mathcal{S}',l'},\bm{\lambda}_{\mathcal{S}',l',n}), \\
             0, ~ \text{otherwise},
             \end{cases}\label{general_mu}\\
&c_{\mathcal{S},l,n}(\mathbf{h},\gamma_{\mathcal{S},l},\bm{\lambda}_{\mathcal{S},l,n})\nonumber\\
&=\mu_{\mathcal{S},l,n}(\mathbf{h},\gamma_{\mathcal{S},l},\bm{\lambda}_{\mathcal{S},l,n})B\left[\log_2\frac{\gamma_{\mathcal{S},l}}{\ln2\sum\nolimits_{k\in\mathcal{S}}\lambda_{\mathcal{S},l,n,k}}\right]^{+},\label{general_c}\\
&\mathbf{W}_{\mathcal{S},l,n}(\mathbf{h},\gamma_{\mathcal{S},l},\bm{\lambda}_{\mathcal{S},l,n})\nonumber\\
&=\frac{\mu_{\mathcal{S},l,n}( \mathbf{h},\gamma_{\mathcal{S},l},\bm{\lambda}_{\mathcal{S},l,n})\mathbf{A}_{\mathcal{S},l,n}^{(t-1)}
\sum\limits_{k\in\mathcal{S}}\lambda_{\mathcal{S},l,n,k}\beta_k|\mathbf{h}_{n,k}^H\mathbf{W}^{(t-1)}_{\mathcal{S},l,n}(\mathbf{h}) |^2}
{\|\mathbf{A}_{\mathcal{S},l,n}^{(t-1)}\|_2^2}.\label{general_w}
\end{align}}

\noindent\normalsize{where}
\begin{equation}
\mathbf{A}_{\mathcal{S},l,n}^{(t-1)}=\sum\nolimits_{k\in\mathcal{S}}\lambda_{\mathcal{S},l,n,k}\beta_k(\mathbf{W}^{(t-1)}_{\mathcal{S},l,n}(\mathbf{h}) )^H\mathbf{h}_{n,k} \mathbf{h}_{n,k}^H.
\end{equation}
Let $\gamma_{\mathcal{S},l}^{\diamond}(\mathbf{h})$ and $\bm{\lambda}_{\mathcal{S},l,n}^{\diamond}(\mathbf{h})$ denote the roots of
\begin{align*}
&\mu_{\mathcal{S},l,n}(\mathbf{h},\gamma_{\mathcal{S},l},\bm{\lambda}_{\mathcal{S},l,n})\left(2^{\frac{c_{\mathcal{S},l,n}(\mathbf{h},\gamma_{\mathcal{S},l},\bm{\lambda}_{\mathcal{S},l,n})}{B\mu_{\mathcal{S},l,n}(\mathbf{h},\gamma_{\mathcal{S},l},\bm{\lambda}_{\mathcal{S},l,n})}}-1\right)
\nonumber\\
&-\frac{2\beta_kR\left\{(\mathbf{W}^{(t^{\diamond})}_{\mathcal{S},l,n}(\mathbf{h}) )^H\mathbf{h}_{n,k} \mathbf{h}_{n,k}^H\mathbf{W}_{\mathcal{S},l,n}(\mathbf{h},\gamma_{\mathcal{S},l},\bm{\lambda}_{\mathcal{S},l,n})\right\}}{M\sigma^2}\nonumber\\
&+\frac{|\mathbf{h}_{n,k}^H\mathbf{W}^{(t^{\diamond})}_{\mathcal{S},l,n}(\mathbf{h}) |^2}{M\sigma^2}=0,~\mathcal{S}\in\mathcal{I},l\in\mathcal{L}_{\mathcal{S}},k\in\mathcal{K}_{\mathcal{S},l},n\in\mathcal{N},\\
&\sum\nolimits_{n\in\mathcal{N}}c_{\mathcal{S},l,n}(\mathbf{h},\gamma_{\mathcal{S},l},\bm{\lambda}_{\mathcal{S},l,n})=|\mathcal{P}_{\mathcal{S}}|D_l,~\mathcal{S}\in\mathcal{I},l\in\mathcal{L}_{\mathcal{S}}.
\end{align*}
An optimal solution of Problem~\ref{CRP1} for the $t^{\diamond}$-th iteration which provides binary subcarrier assignment is given below.
\begin{Cla}[Optimal Solution of Problem~\ref{CRP1}
 for $t^{\diamond}$]\label{lem:opt-ofdma2}
Suppose that  there exists a unique pair $(\mathcal{S}_n,l_n)$ such that
\begin{align}
&G_{\mathcal{S}_n,l_n,n}(\mathbf{h},\gamma_{\mathcal{S}_n,l_n}^{\diamond}(\mathbf{h}),\bm{\lambda}_{\mathcal{S}_n,l_n,n}^{\diamond}(\mathbf{h}))\nonumber\\
&=\max\nolimits_{\mathcal{S}\in\mathcal{I},l\in\mathcal{L}_{\mathcal{S}}}
G_{\mathcal{S},l,n}(\mathbf{h},\gamma_{\mathcal{S},l}^{\diamond}(\mathbf{h}),\bm{\lambda}_{\mathcal{S},l,n}^{\diamond}(\mathbf{h})),
n\in\mathcal{N},\nonumber
\end{align}
Then,
an optimal solution of Problem~\ref{CRP1}  for $t^{\diamond}$
 is given by
 $\mathbf{W}_{\mathcal{S},l,n}^{\diamond}(\mathbf{h}) =\mathbf{W}_{\mathcal{S},l,n}(\mathbf{h},\gamma_{\mathcal{S},l}^{\diamond}(\mathbf{h}),\bm{\lambda}_{\mathcal{S},l,n}^{\diamond}(\mathbf{h}))$,
 $\mu_{\mathcal{S},l,n}^{\diamond}(\mathbf{h})  =\mu_{\mathcal{S},l,n}(\mathbf{h},\gamma_{\mathcal{S},l}^{\diamond}(\mathbf{h}),\bm{\lambda}_{\mathcal{S},l,n}^{\diamond}(\mathbf{h}))$
 and
 $c_{\mathcal{S},l,n}^{\diamond}(\mathbf{h})  =c_{\mathcal{S},l,n}(\mathbf{h},\gamma_{\mathcal{S},l}^{\diamond}(\mathbf{h}),\bm{\lambda}_{\mathcal{S},l,n}^{\diamond}(\mathbf{h}))$.
\end{Cla}

Similar to the condition stated in Claim~1,  the condition in Claim~\ref{lem:opt-ofdma2} can be easily satisfied. Note that $\gamma_{\mathcal{S},l}^{\diamond}(\mathbf{h})$ and $\bm{\lambda}_{\mathcal{S},l,n}^{\diamond}(\mathbf{h})$ can be obtained using a subgradient method.
The details for obtaining a suboptimal solution $(\bm{\mu}^{\diamond}(\mathbf{h}),\bm{\eta}^{\diamond}(\mathbf{h}),\mathbf{c}^{\diamond}(\mathbf{h}),\mathbf{w}^{\diamond}(\mathbf{h}))$
of Problem~1 are summarized in Algorithm~2. Specifically, in Steps 1-5, we solve Problem~4 with computational complexity $\mathcal{O}(M^{3}K^{4}N^{3.5}I^{3.5})$; in Steps 6-13, we obtain an optimal solution of Problem~5 based on the optimal solution of Problem~4 with computational complexity  $\mathcal{O}(M^{5}NK^{2}I)$; in Steps 14-15, we compute a suboptimal solution of Problem~1 based on the optimal solution of Problem~5 with computational complexity $\mathcal{O}(MNKI)$. Therefore, the computational complexity of Algorithm~2 is $\mathcal{O}(M^{3}K^{4}N^{3.5}I^{3.5})$.

\small{\begin{algorithm}
    \caption{\small{Suboptimal Solution of Problem~\ref{P1} for the General Case}}
\begin{footnotesize}
     \begin{algorithmic}[1]
     \STATE Find a random feasible point of Problem~\ref{RP1} as the initial point $(\mathbf{W}^{(0)}(\mathbf{h}),\bm{\mu}^{(0)}(\mathbf{h}),\mathbf{c}^{(0)}(\mathbf{h}))$, and set $t=0$;
     \REPEAT
     \STATE Set $t=t+1$;
     \STATE Obtain $(\mathbf{W}^{(t)}(\mathbf{h}),\bm{\mu}^{(t)}(\mathbf{h}),\mathbf{c}^{(t)}(\mathbf{h}))$
     by  solving Problem~\ref{CRP1} using standard convex optimization techniques;
     \UNTIL{convergence criteria are met}
     \STATE Set $t^{\diamond}=t$;
     \STATE Initialize $\bm{\gamma}^{(1)}$ and $\bm{\lambda}^{(1)}$, and set $i=0$;
     \REPEAT
     \STATE Set $i=i+1$;
           \STATE For all $\mathcal{S}\in\mathcal{I},l\in\mathcal{L}_{\mathcal{S}}$ and $n\in\mathcal{N}$, compute $G_{\mathcal{S},l,n}(\mathbf{h},\gamma_{\mathcal{S},l}^{(i)},\bm{\lambda}_{\mathcal{S},l,n}^{(i)})$, $\mu_{\mathcal{S},l,n}(\mathbf{h},\gamma_{\mathcal{S},l}^{(i)},\bm{\lambda}_{\mathcal{S},l,n}^{(i)})$,
           $c_{\mathcal{S},l,n}(\mathbf{h},\gamma_{\mathcal{S},l}^{(i)},\bm{\lambda}_{\mathcal{S},l,n}^{(i)})$  and $\mathbf{W}_{\mathcal{S},l,n}(\mathbf{h},\gamma_{\mathcal{S},l}^{(i)},\bm{\lambda}_{\mathcal{S},l,n}^{(i)})$  according to \eqref{cst:dc-g}, \eqref{general_mu}, \eqref{general_c} and
           \eqref{general_w}, respectively;
           \STATE For all $\mathcal{S}\in\mathcal{I}$, $l\in\mathcal{L}_{\mathcal{S}}$, $n\in\mathcal{N}$ and $k\in\mathcal{K}_{\mathcal{S},l}$, compute $\lambda^{(i+1)}_{\mathcal{S},l,n,k}$ according to \eqref{eqn-alg2},
where \eqref{eqn-alg2} is shown at the top of the next page, and $\delta^{(i)}$, $i=1,2,\ldots$ satisfy \eqref{cst:step3};
\STATE For all $\mathcal{S}\in\mathcal{I}$ and $l\in\mathcal{L}_{\mathcal{S}}$, compute $\gamma^{(i+1)}_{\mathcal{S},l}$ according to:
\begin{equation}
\gamma_{\mathcal{S},l}^{(i+1)}=\left[\gamma_{\mathcal{S},l}^{(i)}-\delta^{(i)}
\left(\sum\limits_{n\in\mathcal{N}}c_{\mathcal{S},l,n}(\mathbf{h},\gamma_{\mathcal{S},l}^{(i)},\bm{\lambda}_{\mathcal{S},l,n}^{(i)})-|\mathcal{P}_{\mathcal{S}}|D_l\right)\right]^+,
\nonumber
\end{equation}
where $\delta^{(i)}$, $i=1,2,\ldots$
 satisfy \eqref{cst:step3};
     \UNTIL{convergence criteria are met}
     \STATE Set $\bm{\gamma}^{\diamond}(\mathbf{h})=\bm{\gamma}^{(i)}$ and $\bm{\lambda}^{\diamond}(\mathbf{h})=\bm{\lambda}^{(i)}$;
     \STATE For all $\mathcal{S}\in\mathcal{I},~l\in\mathcal{L}_{\mathcal{S}}$ and $n\in\mathcal{N}$, set
     $\mu_{\mathcal{S},l,n}^{\diamond}(\mathbf{h})=\mu_{\mathcal{S},l,n}(\mathbf{h},\gamma_{\mathcal{S},l}^{\diamond}(\mathbf{h}),\bm{\lambda}_{\mathcal{S},l,n}^{\diamond}(\mathbf{h}))$,
      $\eta_{\mathcal{S},l,n}^{\diamond}(\mathbf{h})=\|\mathbf{W}_{\mathcal{S},l,n}(\mathbf{h},\gamma_{\mathcal{S},l}^{\diamond}(\mathbf{h}),\bm{\lambda}_{\mathcal{S},l,n}^{\diamond}(\mathbf{h}))\|_2$, $c_{\mathcal{S},l,n}^{\diamond}(\mathbf{h})=c_{\mathcal{S},l,n}(\mathbf{h},\gamma_{\mathcal{S},l}^{\diamond}(\mathbf{h}),\bm{\lambda}_{\mathcal{S},l,n}^{\diamond}(\mathbf{h}))$
      and $\mathbf{w}_{n}^{\diamond}(\mathbf{h})=\sum\nolimits_{\mathcal{S}\in\mathcal{I}}\sum\nolimits_{l\in\mathcal{L}_{\mathcal{S}}}\mu_{\mathcal{S},l,n}^{\diamond}(\mathbf{h})\frac{\mathbf{W}_{\mathcal{S},l,n}(\mathbf{h},\gamma_{\mathcal{S},l}^{\diamond}(\mathbf{h}),\bm{\lambda}_{\mathcal{S},l,n}^{\diamond}(\mathbf{h}))}{\eta_{\mathcal{S},l,n}^{\diamond}(\mathbf{h})}$.
    \end{algorithmic}
    \end{footnotesize}\label{alg:opt-SP1}
\end{algorithm}}


\begin{figure*}[!t]
\small{
\begin{align}
\lambda_{\mathcal{S},l,n,k}^{(i+1)}=&\left[\lambda_{\mathcal{S},l,n,k}^{(i)}-\delta^{(i)}
\left(\mu_{\mathcal{S},l,n}(\mathbf{h},\gamma^{(i)}_{\mathcal{S},l},\bm{\lambda}^{(i)}_{\mathcal{S},l,n})\left(2^{\frac{c_{\mathcal{S},l,n}(\mathbf{h},\gamma^{(i)}_{\mathcal{S},l},\bm{\lambda}^{(i)}_{\mathcal{S},l,n})}{B\mu_{\mathcal{S},l,n}(\mathbf{h},\gamma^{(i)}_{\mathcal{S},l},\bm{\lambda}^{(i)}_{\mathcal{S},l,n})}}-1\right)
\right.\right.\nonumber\\
&\left.\left.-\frac{2\beta_kR\left\{(\mathbf{W}^{(t^{\diamond})}_{\mathcal{S},l,n}(\mathbf{h}))^H\mathbf{h}_{n,k}\mathbf{h}_{n,k}^H\mathbf{W}_{\mathcal{S},l,n}(\mathbf{h},\gamma^{(i)}_{\mathcal{S},l},\bm{\lambda}^{(i)}_{\mathcal{S},l,n})\right\}}{M\sigma^2}\right.\right.\left.\left.+\frac{|\mathbf{h}_{n,k}^H\mathbf{W}^{(t^{\diamond})}_{\mathcal{S},l,n}(\mathbf{h})|^2}{M\sigma^2}\right)\right]^+,
\label{eqn-alg2}
\end{align}}
\hrulefill
\end{figure*}

\section{Average Transmission Power Minimization With User Transcoding}
\normalsize{In} this section, we consider the case with user transcoding.
Although message $(\mathcal{S},l)$, where $\mathcal{S}\in\mathcal{I}$ and $l\in\mathcal{L}$, is requested only by the users in $\mathcal{K}_{\mathcal{S},l}$,
it may be transmitted to all users in $\mathcal{K}_{\mathcal{S},l}$ and
$\mathcal{K}_{\mathcal{S},l'}$ for some $l'<l$ simultaneously  via multicast.
The users in $\mathcal{K}_{\mathcal{S},l}$ directly use message $(\mathcal{S},l)$.
In contrast, the users in $\mathcal{K}_{\mathcal{S},l'}$, $l'<l$ first convert message
$(\mathcal{S},l)$ to message $(\mathcal{S},l')$ using transcoding, before using it.
This type of multicast opportunities are referred to as transcoding-enabled multicast opportunities \cite{long2}. An  illustration example is shown in Fig.~\ref{fig:system-model} (c).
 In this example,  by making use of natural  multicast opportunities, the server multicasts message $(\{1,2\},1)$ to user 1 and user 2;
 by making use of transcoding-enabled multicast opportunities, the server multicasts
 message $(\{2,3\},2)$ to user 2 and user 3  and multicasts
 message $(\{1,2,3\},2)$ to user 1, user 2  and  user 3.
By comparing Fig.~\ref{fig:system-model} (b) and Fig.~\ref{fig:system-model} (c), we can see that transcoding provides  more multicast opportunities.

To model user transcoding \cite{long-tc},
let $\mathbf{x}\triangleq(x_{\mathcal{S},l,k})_{\mathcal{S}\in\mathcal{I},l\in\mathcal{L},k\in\mathcal{S}}$
denote the quality level selection variables,
where
\begin{align}
&x_{\mathcal{S},l,k}\in\{0,1\},~\mathcal{S}\in\mathcal{I},l\in\mathcal{L},k\in\mathcal{S},\label{cst:x}\\
&\sum\nolimits_{l\in\mathcal{L}}x_{\mathcal{S},l,k}=1,~\mathcal{S}\in\mathcal{I},k\in\mathcal{S}.\label{cst:sumx}
\end{align}
Here, $x_{\mathcal{S},l,k}=1$ indicates that the server will transmit message $(\mathcal{S},l)$ to user $k$, and $x_{\mathcal{S},l,k}=0$ otherwise.
 Note that the constraints in \eqref{cst:sumx} ensure that the server transmits only one
 of the messages $(\mathcal{S},l)$, $l\in\mathcal{L}$ which has quality level
 $\sum\nolimits_{l'\in\mathcal{L}}l'x_{\mathcal{S},l',k}$ to user $k$. Note that
 $\mathbf{x}$ should not change during the considered time duration because of the video coding structure.
With transcoding, to ensure that user $k$ can play his FoV at quality level $r_k$, it is sufficient to require:
\begin{equation}
\sum\nolimits_{l\in\mathcal{L}}lx_{\mathcal{S},l,k}\geq r_k,~\mathcal{S}\in\mathcal{I},l\in\mathcal{L},k\in\mathcal{S}.\label{cst:sumlx}
\end{equation}
Then, the successful transmission constraints in  \eqref{cst:minrate}
become:
\begin{align}
     &\mu_{\mathcal{S},l,n}(\mathbf{h})B\log_2\left(1+
    \frac{\eta_{\mathcal{S},l,n}(\mathbf{h})\beta_{k}|\mathbf{h}_{n,k}^H\mathbf{w}_{n}(\mathbf{h})|^2}
    {M\sigma^2}\right)\nonumber\\
    &\geq c_{\mathcal{S},l,n}(\mathbf{h})x_{\mathcal{S},l,k},\mathcal{S}\in\mathcal{I},l\in\mathcal{L},k\in\mathcal{S},n\in\mathcal{N},\label{cst:minrate-p2}
\end{align}
and
the transmission rate constraints in
\eqref{cst:averate} become:
\begin{equation}
\sum\limits_{n\in\mathcal{N}}c_{\mathcal{S},l,n}(\mathbf{h})\geq |\mathcal{P}_{\mathcal{S}}|D_{l}x_{\mathcal{S},l,k},~\mathcal{S}\in\mathcal{I},l\in\mathcal{L},k\in\mathcal{S}.\label{cst:averate-p2}
\end{equation}

On the other hand, user transcoding also consumes power. For ease of exposition, we assume that the transcoding powers for reducing the quality levels of all tiles by one are identical at each user. Denote $E_k$ as the transcoding power at user $k$ for lowering the quality level of the representation of a tile by one. Since different users have heterogeneous
hardware conditions, we allow $E_k$, $k\in\mathcal{K}$ to be different.
Then, the total transcoding power  at all users is
$E_{\text{tc}}(\mathbf{x})\triangleq\sum\nolimits_{\mathcal{S}\in\mathcal{I}}\sum\nolimits_{k\in\mathcal{S}}\left(\sum\nolimits_{l\in\mathcal{L}}lx_{\mathcal{S},l,k}-r_k\right)|\mathcal{P}_{\mathcal{S}}|E_k$.
The weighted sum~of the average transmission power and the transcoding
power is
\begin{equation}
\mathbb{E}\left[\frac{1}{M}\sum\limits_{n\in\mathcal{N}}\sum\limits_{\mathcal{S}\in\mathcal{I}}\sum\limits_{l\in\mathcal{L}}\mu_{\mathcal{S},l,n}(\mathbf{H})\eta_{\mathcal{S},l,n}(\mathbf{H})\right] +\alpha E_{\text{tc}}(\mathbf{x}),\nonumber
\end{equation}
where $\alpha\geq1$ is the corresponding weight factor, and
the expectation is taken over $\mathbf{H} $.
Note that $\alpha>1$ means that a higher cost  on the power consumption for user devices is incurred due to the limited battery powers of user devices.


With slight abuse of notation,
denote
$\bm{\mu}(\mathbf{h})\triangleq(\mu_{\mathcal{S},l,n}(\mathbf{h}))_{\mathcal{S}\in\mathcal{I},l\in\mathcal{L},n\in\mathcal{N}}$,
$\bm{\eta}(\mathbf{h})\triangleq(\eta_{\mathcal{S},l,n}(\mathbf{h}))_{\mathcal{S}\in\mathcal{I},l\in\mathcal{L},n\in\mathcal{N}}$, $\mathbf{c}(\mathbf{h})\triangleq(c_{\mathcal{S},l,n}(\mathbf{h}))_{\mathcal{S}\in\mathcal{I},l\in\mathcal{L},n\in\mathcal{N}}$
and $\mathbf{w}(\mathbf{h})\triangleq(\mathbf{w}_{n}(\mathbf{h}))_{n\in\mathcal{N}}$.
Similarly, we treat $\bm{\mu}(\mathbf{h}),\bm{\eta}(\mathbf{h}),\mathbf{c}(\mathbf{h}),\mathbf{w}(\mathbf{h})$ as functions of $\mathbf{h}$, respectively.
For given quality requirements of all users $\mathbf{r}$, we would like to  minimize the weighted sum of the average transmission power and the transcoding power under the constraints in
\eqref{cst:mu1}-\eqref{cst:w}, \eqref{cst:c}, and \eqref{cst:x}-\eqref{cst:averate-p2}, by optimizing  $\bm{\mu}(\mathbf{h})$,  $\bm{\eta}(\mathbf{h})$, $\mathbf{c}(\mathbf{h})$, $\mathbf{w}(\mathbf{h})$, and  $\mathbf{x}$. Specifically, for given $\mathbf{r}$, we have
\begin{Prob}[Average Total Transmission Power and \linebreak Transcoding Power Minimization]\label{P2}
\begin{align}
 \min\limits_{\bm{\mu}(\mathbf{h}),\bm{\eta}(\mathbf{h}),\mathbf{c}(\mathbf{h}),\mathbf{w}(\mathbf{h}),\mathbf{x}}~&\mathbb{E}\left[\frac{1}{M}\sum\limits_{n\in\mathcal{N}}\sum\limits_{\mathcal{S}\in\mathcal{I}}\sum\limits_{l\in\mathcal{L}}\mu_{\mathcal{S},l,n}(\mathbf{H})\eta_{\mathcal{S},l,n}(\mathbf{H})\right] \nonumber\\&+\alpha E_{\text{tc}}(\mathbf{x}) \nonumber\\
    \mathrm{s.t.} ~~
    \eqref{cst:mu1},~\eqref{cst:mu2},~\eqref{cst:eta},~&\eqref{cst:w},~\eqref{cst:c}, ~\eqref{cst:x},~\eqref{cst:sumx},~\eqref{cst:sumlx},~\eqref{cst:minrate-p2},~\eqref{cst:averate-p2}.\nonumber
\end{align}
\end{Prob}

Problem~\ref{P2} is a variational problem. Besides, it is a
 two-timescale  mixed optimization problem, and is more challenging
 than Problem~\ref{P1}.\footnote{Similarly, it is in general impossible to analytically or numerically show the gap between a globally optimal solution and a suboptimal solution\cite{NL}.}
Specifically, quality level selection is in a larger timescale and adapts to the
channel distribution; subcarrier, power, and rate allocation and beamforming design are
in a shorter timescale and are adaptive to instantaneous channel states.\footnote{The optimal quality level selection can be used until $\mathcal{G}_{k},k\in\mathcal{K}$ change. For any given $\mathcal{G}_{k},k\in\mathcal{K}$, we only need to solve Problem~\ref{P2} once and then solve Problem~8 (which is similar to Problem 1) for each subsequent frame.} Problem~\ref{P2} generalizes multi-group multicast problems in MIMO-OFDMA systems because it allows optimizing multicast groups to a certain extent via exploiting transcoding-enabled multicast opportunities. In the following,  Problem \ref{P2} is solved using two methods. The first method provides optimal solutions for some special cases, and the second method offers a suboptimal solution for the general~case.



\subsection{Solutions for Special Cases}\label{4a}


It can be easily veryfied that an optimal solution satisfies
\begin{equation}
\sum\nolimits_{l\in\mathcal{L}}lx_{\mathcal{S},l,k} \in \mathcal{L}_{\mathcal{S}},~\mathcal{S}\in\mathcal{I},~l\in\mathcal{L},~k\in\mathcal{S}.\label{new}
\end{equation}
Thus, we can impose the extra constraints in \eqref{new} without loss of optimality.
In two special cases,
we solve Problem~\ref{P2} by solving the following equivalent problem of Problem~\ref{P2}.


\begin{Prob}[Equivalent Problem of Problem~\ref{P2}]\label{EXP2}
\begin{equation}
\min\nolimits_{\mathbf{x}\in\bm{\mathcal{X}}}~\mathbb{E}\left[E^{\star}(\mathbf{x},\mathbf{H})\right]+\alpha E_{\text{tc}}(\mathbf{x})\nonumber
\end{equation}
where $\bm{\mathcal{X}}\triangleq\{\mathbf{x}|\mathbf{x}~\text{satisfies}~\eqref{cst:x},~\eqref{cst:sumx},~\eqref{cst:sumlx},~\eqref{new}\}$, and $E^{\star}(\mathbf{x},\mathbf{h})$ is given by the following problem.
\end{Prob}
\begin{Prob}[Subproblem of Problem~\ref{EXP2} for
$\mathbf{x}\in\bm{\mathcal{X}}$ and $\mathbf{h} $]\label{EP2}
\begin{align}
E^{\star}(\mathbf{x},\mathbf{h})\triangleq&\min_{\mathbf{w}(\mathbf{h}),\bm{\mu}(\mathbf{h}),\bm{\eta}(\mathbf{h}),\mathbf{c}(\mathbf{h})}~
\sum\limits_{n\in\mathcal{N}}\sum\limits_{\mathcal{S}\in\mathcal{I}}\sum\limits_{l\in\mathcal{L}}
\frac{\mu_{\mathcal{S},l,n}(\mathbf{h})\eta_{\mathcal{S},l,n}(\mathbf{h})}{M}\nonumber\\
    &\mathrm{s.t.} ~~
    \eqref{cst:mu1},~\eqref{cst:mu2},~\eqref{cst:eta},~\eqref{cst:w},~\eqref{cst:c},~\eqref{cst:minrate-p2},~\eqref{cst:averate-p2}.\nonumber
\end{align}
\end{Prob}


Problem~\ref{EP2} has the same structure as Problem~\ref{P1} and can be equivalently converted to the following problem.

\begin{Prob}[Equivalent Problem of Problem~\ref{EP2} for $\mathbf{x}\in\bm{\mathcal{X}}$ and $\mathbf{h}$]\label{p8equal}
\begin{align}
&\min_{\bm{\mu}(\mathbf{h}),\mathbf{P}(\mathbf{h}) }~\frac{1}{M}\sum\nolimits_{n\in\mathcal{N}}\sum\nolimits_{\mathcal{S}\in\mathcal{I}}\sum\nolimits_{l\in\mathcal{L}_{\mathcal{S}}}P_{\mathcal{S},l,n}(\mathbf{h}) \nonumber\\
    &\mathrm{s.t.} ~~
    \eqref{cst:mu1},~\eqref{cst:mu2},~\eqref{cst:p},\nonumber\\
&\sum\nolimits_{n\in\mathcal{N}}\mu_{\mathcal{S},l,n}(\mathbf{h})B\log_2\left(1+
    \frac{P_{\mathcal{S},l,n}(\mathbf{h})}
    {\mu_{\mathcal{S},l,n}(\mathbf{h})Q_{\mathcal{S},l,n}^{\dagger}(\mathbf{h})}\right)\nonumber\\
    &\geq |\mathcal{P}_{\mathcal{S}}|D_{l}x_{\mathcal{S},l,k},~\mathcal{S}\in\mathcal{I},l\in\mathcal{L}_{\mathcal{S}},k\in\mathcal{K}_{\mathcal{S},l},\label{cst:sumraten}
\end{align}
where $Q_{\mathcal{S},l,n}^{\dagger}(\mathbf{h})$ is given by Problem \ref{BF1}. Let $(\hat{\bm{\mu}}^{\dagger}(\mathbf{h}), \hat{\mathbf{P}}^{\dagger}(\mathbf{h}))$ denote an optimal solution of Problem \ref{p8equal}.
\end{Prob}

Analogously, by exploring structures of Problem \ref{EP2}, Problem \ref{BF1}, and Problem \ref{p8equal}, we have the following result.
\begin{Thm}[Equivalence between Problem~\ref{EP2} and Problem~\ref{p8equal}]
The optimal values of Problem~\ref{EP2} and Problem~\ref{p8equal} are identical.
In addition, $(\hat{\bm{\mu}}^{\dagger}(\mathbf{h}),\hat{\bm{\eta}}^{\dagger}(\mathbf{h}),\hat{\mathbf{c}}^{\dagger}(\mathbf{h}),
\hat{\mathbf{w}}^{\dagger}(\mathbf{h}))$
is an optimal solution of Problem~\ref{EP2}, where $\hat{\bm{\eta}}^{\dagger}(\mathbf{h})=\hat{\mathbf{P}}^{\dagger}(\mathbf{h})$, $\hat{\mathbf{w}}^{\dagger}_{n}(\mathbf{h}) =\sum\nolimits_{\mathcal{S}\in\mathcal{I}}\sum\nolimits_{l\in\mathcal{L}_{\mathcal{S}}}\hat{\mu}_{\mathcal{S},l,n}^{\dagger}(\mathbf{h})\frac{\mathbf{v}^{\dagger}_{\mathcal{S},l,n}(\mathbf{h}) }{\sqrt{Q_{\mathcal{S},l,n}^{\dagger}(\mathbf{h})}}$, and $\hat{\mathbf{c}}^{\dagger}(\mathbf{h})\triangleq(\hat{c}^{\dagger}_{\mathcal{S},l,n}(\mathbf{h}))_{\mathbf{S}\in\mathcal{I},l\in\mathcal{L}_{\mathcal{S}},n\in\mathcal{N}}$
with $\hat{c}_{\mathcal{S},l,n}^{\dagger}(\mathbf{h}) =\hat{\mu}_{\mathcal{S},l,n}^{\dagger}(\mathbf{h})B\log_2\left(1+
    \frac{\hat{P}_{\mathcal{S},l,n}^{\dagger}(\mathbf{h}) }
    {Q_{\mathcal{S},l,n}^{\dagger}(\mathbf{h}) }\right)$.
\end{Thm}

According to Theorem 3, to obtain an optimal solution of Problem \ref{EP2}, we can first obtain $\hat{\mathbf{w}}^{\dagger}(\mathbf{h})$ by solving Problem \ref{BF1}, and then obtain $\hat{\bm{\mu}}^{\dagger}(\mathbf{h}),\hat{\bm{\eta}}^{\dagger}(\mathbf{h}),$ and $\hat{\mathbf{c}}^{\dagger}(\mathbf{h})$ by solving Problem \ref{p8equal}.
In the case where each group $\mathcal{S}\in\mathcal{I}$ has at most 3 users, i.e., $|\mathcal{S}|\leq3,~\mathcal{S}\in\mathcal{I}$, we can obtain an optimal solution of Problem~\ref{p8equal} by using Algorithm~\ref{alg:opt-GP1} in Section~\ref{sec:p1-sp1}.
In the case of a large antenna array, we can get an~asymptotically optimal solution of Problem~\ref{p8equal} using the method in Section~\ref{sect:p1-sp2}.
After obtaining $E^{\star}(\mathbf{x},\mathbf{h})$ for all $\mathbf{x}\in\bm{\mathcal{X}}$ and
$\mathbf{h} $, we can numerically compute $\mathbb{E}[E^{\star}(\mathbf{x},\mathbf{H})]$, for all $\mathbf{x}\in\bm{\mathcal{X}}$, and then solve Problem~\ref{EXP2} using the exhaustive search. The exhaustive search is over $\prod\limits_{\mathcal{S}\in\mathcal{I}}L_{\mathcal{S}}$ possible values of $\mathbf{x}$, i.e., the computational complexity scales with $\prod\limits_{\mathcal{S}\in\mathcal{I}}L_{\mathcal{S}}$, where $L_{\mathcal{S}} \triangleq |\mathcal{L}_{\mathcal{S}}|$.\footnote{When $\prod_{\mathcal{S}\in\mathcal{I}}L_{\mathcal{S}}$ is large, one can adopt Algorithm 2 to obtain a suboptimal solution with relatively low computational complexity.}

%


\subsection{Suboptimal Solution in General Case} \label{4b}
In the general case,  we directly tackle Problem~\ref{P2} and develop a low-complexity
algorithm to obtain a suboptimal
solution.
Specifically, we get a suboptimal quality level selection by solving an approximation of Problem~\ref{P2} using DC programming and obtain a suboptimal  solution of Problem~\ref{EP2} with the obtained $\mathbf{x}$ using Algorithm~\ref{alg:opt-SP1} in Section~\ref{sec:p1-dc}.

First, we  get an approximation of Problem~\ref{P2},
which has only one timescale and has a much smaller number of variables than
Problem~\ref{P2} \cite{xu_tcom}. Specifically,
\eqref{cst:averate-p2} and \eqref{cst:minrate-p2} are replaced by
\begin{align}
&\sum\nolimits_{n\in\mathcal{N}}\bar{c}_{\mathcal{S},l,n}\geq |\mathcal{P}_{\mathcal{S}}|D_{l}x_{\mathcal{S},l,k},~\mathcal{S}\in\mathcal{I},l\in\mathcal{L},k\in\mathcal{S},\label{cst:averate-ap2ex}\\
&\bar{\mu}_{\mathcal{S},l,n}B\log_2\left(1+
    \frac{\bar{\eta}_{\mathcal{S},l,n}}
    {\bar{Q}_{k}}\right)\geq \bar{c}_{\mathcal{S},l,n}x_{\mathcal{S},l,k},\nonumber\\
&\mathcal{S}\in\mathcal{I},l\in\mathcal{L},k\in\mathcal{S},n\in\mathcal{N},\label{cst:averate-ap2ex2}
\end{align}
respectively.
Here, $\bar{\mu}_{\mathcal{S},l,n}$, $\bar{\eta}_{\mathcal{S},l,n}$ and $\bar{c}_{\mathcal{S},l,n}$
approximately characterize $\mathbb{E}[\mu_{\mathcal{S},l,n}(\mathbf{H})]$,
$\mathbb{E}[\eta_{\mathcal{S},l,n}(\mathbf{H})]$ and
$\mathbb{E}[c_{\mathcal{S},l,n}(\mathbf{H})]$, and
\begin{align}
&\bar{Q}_{k}\triangleq\mathbb{E}\left[\min_{\mathbf{w}_{n}(\mathbf{H})\in\{\mathbf{w}|\mathbf{w}\in\mathbb{C}^{M\times1},\|\mathbf{w}\|_2=1\}}\frac{M\sigma^2}{\beta_{k}|\mathbf{h}_{n,k}^H\mathbf{w}_{n}(\mathbf{H})|^2}\right]\nonumber\\
&=\frac{M\sigma^2}{\beta_k\mathbb{E}\left[\max\nolimits_{\mathbf{w}_{n}(\mathbf{H})\in\{\mathbf{w}|\mathbf{w}\in\mathbb{C}^{M\times1},\|\mathbf{w}\|_2=1\}}|\mathbf{h}_{n,k}^H\mathbf{w}_{n}(\mathbf{H})|^2\right]}
\nonumber\\
&=\frac{\sigma^2}{\beta_k}.\nonumber
\end{align}

\noindent \normalsize{Then, by introducing $\bar{p}_{\mathcal{S},l,n}\triangleq\bar{\mu}_{\mathcal{S},l,n}\bar{\eta}_{\mathcal{S},l,n}$ and eliminating $\bar{\mathbf{c}}$ as well as $\bar{\bm{\eta}}$, we
simplify the constraints in
\eqref{cst:averate-ap2ex} and \eqref{cst:averate-ap2ex2} to}
\begin{align}
    &\sum\nolimits_{n\in\mathcal{N}}\bar{\mu}_{\mathcal{S},l,n}B\log_2\left(1+
    \frac{\bar{p}_{\mathcal{S},l,n}}
    {\bar{Q}_{k}\bar{\mu}_{\mathcal{S},l,n}}\right)\geq |\mathcal{P}_{\mathcal{S}}|D_lx_{\mathcal{S},l,k},\nonumber\\
    &\mathcal{S}\in\mathcal{I},l\in\mathcal{L},k\in\mathcal{S},n\in\mathcal{N}.\label{cst:minrate2}
\end{align}
Therefore, we can obtain the following problem.
\begin{Prob}[Approximation of Problem~\ref{P2}]\label{AAP2}
\begin{align}
 E_t^{\dagger}\triangleq\min_{\bar{\bm{\mu}},\bar{\mathbf{p}},\mathbf{x}}~&\frac{1}{M}\sum\limits_{n\in\mathcal{N}}\sum\limits_{\mathcal{S}\in\mathcal{I}}\sum\limits_{l\in\mathcal{L}}\bar{p}_{\mathcal{S},l,n}
 +\alpha E_{\text{tc}}(\mathbf{x})\nonumber\\
    \mathrm{s.t.} ~~&\eqref{cst:x},~\eqref{cst:sumx},~\eqref{cst:sumlx},~\eqref{cst:minrate2},\nonumber\\
     &\bar{\mu}_{\mathcal{S},l,n}\geq0,~\mathcal{S}\in\mathcal{I},l\in\mathcal{L},n\in\mathcal{N},\label{cst:ap2-mu}\\
     &\bar{p}_{\mathcal{S},l,n}\geq0,~\mathcal{S}\in\mathcal{I},l\in\mathcal{L},n\in\mathcal{N},\label{cst:ap2-eta}\\
     &\sum\nolimits_{\mathcal{S}\in\mathcal{I}}\sum\nolimits_{l\in\mathcal{L}}\bar{\mu}_{\mathcal{S},l,n}=1,n\in\mathcal{N},\label{cst:ap2-sumu}
\end{align}
where $\bar{\bm{\mu}}\triangleq(\bar{\mu}_{\mathcal{S},l,n})_{\mathcal{S}\in\mathcal{I},l\in\mathcal{L},n\in\mathcal{N}}$ and $\bar{\bm{p}}\triangleq(\bar{\eta}_{\mathcal{S},l,n})_{\mathcal{S}\in\mathcal{I},l\in\mathcal{L},n\in\mathcal{N}}$.
Let $(\bar{\bm{\mu}}^{\dagger},\bar{\bm{p}}^{\dagger},\mathbf{x}^{\dagger})$ denote an optimal solution.
\end{Prob}

Problem~\ref{AAP2} is a single timescale mixed discrete-continuous problem with $LI(2N+K)$ variables,  much simpler than Problem~\ref{P2}. The dimensions of $\bar{\bm{\mu}}$ and $\bar{\bm{p}}$ are both $LNI$.
To further reduce the computational complexity of Problem~\ref{AAP2}, we convert it to an equivalent problem.
\begin{Prob}[Equivalent Problem of Problem~\ref{AAP2}]\label{EAP2}
\begin{align}
 \bar{E}_t^{\ast}\triangleq\min_{\bar{\mathbf{N}},\bar{\mathbf{P}},\mathbf{x}}~&\frac{1}{M}\sum\nolimits_{\mathcal{S}\in\mathcal{I}}\sum\nolimits_{l\in\mathcal{L}}\bar{P}_{\mathcal{S},l}
 +\alpha E_{\text{tc}}(\mathbf{x})\nonumber\\
    \mathrm{s.t.} ~~&\eqref{cst:x},~\eqref{cst:sumx},~\eqref{cst:sumlx},\nonumber\\
     &\bar{N}_{\mathcal{S},l}\geq0,~\mathcal{S}\in\mathcal{I},l\in\mathcal{L},\label{cst:ap2-N}\\
     &\bar{P}_{\mathcal{S},l}\geq0,~\mathcal{S}\in\mathcal{I},l\in\mathcal{L},\label{cst:ap2-P}\\
     &\sum\nolimits_{\mathcal{S}\in\mathcal{I}}\sum\nolimits_{l\in\mathcal{L}}\bar{N}_{\mathcal{S},l}=N,\label{cst:ap2-sumN}\\
     &\bar{N}_{\mathcal{S},l}B\log_2\left(1+
    \frac{\bar{P}_{\mathcal{S},l}}
    {\bar{N}_{\mathcal{S},l}\bar{Q}_{k}}\right)\geq |\mathcal{P}_{\mathcal{S}}|D_lx_{\mathcal{S},l,k},\nonumber\\
    &\mathcal{S}\in\mathcal{I},l\in\mathcal{L},k\in\mathcal{S},n\in\mathcal{N},\label{cst:ap2-rate}
\end{align}
where  $\bar{\mathbf{N}}\triangleq(\bar{N}_{\mathcal{S},l})_{\mathcal{S}\in\mathcal{I},l\in\mathcal{L}}$ and $\bar{\mathbf{P}}\triangleq(\bar{P}_{\mathcal{S},l})_{\mathcal{S}\in\mathcal{I},l\in\mathcal{L}}$.
Let $(\bar{\mathbf{N}}^{\ast},\bar{\mathbf{P}}^{\ast},\mathbf{x}^{\ast})$ denote an optimal solution.
\end{Prob}

\begin{Thm}[Equivalence Between Problem~\ref{AAP2} and Problem~\ref{EAP2}]
There exist an optimal solution of Problem~\ref{AAP2} (i.e., $(\bar{\bm{\mu}}^{\dagger},\bar{\bm{p}}^{\dagger},\mathbf{x}^{\dagger})$) and an optimal solution of Problem~\ref{EAP2} (i.e., $(\bar{\mathbf{N}}^{\ast},\bar{\mathbf{P}}^{\ast},\mathbf{x}^{\ast})$)  such that
$\mathbf{x}^{\ast}=\mathbf{x}^{\dagger}$, and
$\bar{N}_{\mathcal{S},l}^{\ast}=\sum\nolimits_{n\in\mathcal{N}}\bar{\mu}^{\dagger}_{\mathcal{S},l,n}$,
$\bar{P}_{\mathcal{S},l}^{\ast}=\sum\nolimits_{n\in\mathcal{N}}\bar{p}^{\dagger}_{\mathcal{S},l,n}$, $\mathcal{S}\in\mathcal{I}$,~$l\in\mathcal{L}$.
\end{Thm}
\begin{Proof}
 See Appendix~C.
\end{Proof}

By noting that   the dimensions of $\bar{\mathbf{N}}$ and $\bar{\mathbf{P}}$ are both $LI$, i,e, $\frac{1}{N}$ of those of $\bar{\bm{\mu}}$ and $\bar{\bm{\eta}}$, Problem~\ref{EAP2}  is much simpler than Problem~\ref{AAP2}.
In the following,  a low-complexity algorithm is developed to obtain a suboptimal solution of Problem~\ref{EAP2} using the DC algorithm\cite{dc}.

First, we convert Problem~\ref{EAP2} to a penalized DC problem. Specifically, we equivalently convert the discrete constraints in \eqref{cst:x} to the following continuous constraints:
\begin{align}
&0\leq x_{\mathcal{S},l,k}\leq1,~\mathcal{S}\in\mathcal{I},l\in\mathcal{L},k\in\mathcal{S},\label{cst:eqx}\\
&x_{\mathcal{S},l,k}(1-x_{\mathcal{S},l,k})\leq0,~\mathcal{S}\in\mathcal{I},l\in\mathcal{L},k\in\mathcal{S}.\label{cst:eqx2}
\end{align}
By augmenting the constraints in \eqref{cst:eqx2} to the objective function via the penalty method \cite{long}, Problem~\ref{EAP2} can be equivalently converted to the following problem.

\begin{Prob}[Penalized DC Problem of Problem~\ref{EAP2}]\label{DCP2}
\begin{align}
 \min_{\bar{\mathbf{N}},\bar{\mathbf{P}},\mathbf{x}}~&\sum\nolimits_{\mathcal{S}\in\mathcal{I}}\sum\nolimits_{l\in\mathcal{L}}\bar{P}_{\mathcal{S},l}
 +\alpha E_{\text{tc}}(\mathbf{x})+\rho\chi(\mathbf{x})\nonumber\\
    \mathrm{s.t.} ~~
    &\eqref{cst:ap2-N},~\eqref{cst:ap2-P},~\eqref{cst:ap2-sumN},~\eqref{cst:ap2-rate},~\eqref{cst:sumlx},~\eqref{cst:sumx},\nonumber\\
    &0\leq x_{\mathcal{S},l,k}\leq1,~\mathcal{S}\in\mathcal{I},l\in\mathcal{L},k\in\mathcal{S},
\end{align}
\end{Prob}
where the penalty parameter $\rho>0$ and
the penalty function
$\chi(\mathbf{x})\triangleq\sum\nolimits_{\mathcal{S}\in\mathcal{I}}\sum\nolimits_{k\in\mathcal{S}}\sum\nolimits_{l\in\mathcal{L}}
x_{\mathcal{S},l,k}(1-x_{\mathcal{S},l,k})$.

Note that we can regard the objective function of Problem~\ref{DCP2} as a difference of two convex functions and the constraints of Problem~\ref{DCP2}
 are all convex. Thus, we can view Problem~\ref{DCP2}  as a penalized DC problem of Problem~\ref{EAP2}. When the feasible set of Problem~\ref{EAP2} is nonempty, there exists $\rho_0>0$ such that for all $\rho>\rho_0$,
Problem~\ref{DCP2} and Problem~\ref{EAP2} are equivalent \cite{dc}.
 By solving Problem~\ref{DCP2} using
a DC algorithm \cite{dc}, we obtain a stationary point $(\bar{\mathbf{P}}^{\diamond},\bar{\mathbf{N}}^{\diamond},\mathbf{x}^{\diamond})$  of Problem~\ref{EAP2}.

Next, by substituting $\mathbf{x}^{\diamond}$ into Problem~\ref{EP2},    we can get a suboptimal solution of Problem~\ref{EP2} for each $\mathbf{h} $,  denoted by  $(\bm{\mu}^{\diamond}(\mathbf{h}),\bm{\eta}^{\diamond}(\mathbf{h}),\mathbf{c}^{\diamond}(\mathbf{h}),\mathbf{w}^{\diamond}(\mathbf{h}))$, using Algorithm~\ref{alg:opt-SP1} in Section~\ref{sec:p1-dc}.
The details for obtaining a suboptimal solution of Problem~\ref{P2} in the general case, denoted by $(\bm{\mu}^{\diamond},\bm{\eta}^{\diamond},\mathbf{c}^{\diamond},\mathbf{w}^{\diamond},\mathbf{x}^{\diamond})$, are summarized in Algorithm~\ref{alg:DC-P2}.
\begin{algorithm}
    \caption{\small{Suboptimal Solution of Problem~\ref{P2} for the General Case}}
\begin{footnotesize}
     \begin{algorithmic}[1]
\STATE Obtain $\mathbf{x}^{\diamond}$ by solving Problem~\ref{DCP2} using the DC algorithm in \cite{dc};
     \STATE For any $\mathbf{h} $, obtain a suboptimal solution  $(\bm{\mu}^{\diamond}(\mathbf{h}),\bm{\eta}^{\diamond}(\mathbf{h}),\mathbf{c}^{\diamond}(\mathbf{h}),\mathbf{w}^{\diamond}(\mathbf{h}))$ of Problem~\ref{EP2} with $\mathbf{x}=\mathbf{x}^{\diamond}$  using Algorithm~\ref{alg:opt-SP1}.
    \end{algorithmic}
    \end{footnotesize}\label{alg:DC-P2}
\end{algorithm}

\section{Numerical Results}
This section considers the two scenarios  without (w/o) and with (w)
user transcoding, and compares the proposed solutions in Section~III
and Section~IV with baseline schemes.
In the scenario without user transcoding, we consider the following two baseline schemes.
Baseline~1 serves $K$
 users separately (i.e., adopts unicast) and adopts the normalized maximum ratio transmission (MRT) beamformer for each user on each subcarrier.
Baseline~2 jointly considers the FoVs of all users (i.e., adopts multicast for a message, if there exists a multicast opportunity) as in this paper
and adopts the normalized MRT beamformer for a massage on each subcarrier  obtained based on
 the channel power gain matrix of all users requiring this message on each subcarrier \cite{8392787}.
Then, for each baseline scheme,
the optimal subcarrier, power,
and rate allocation is obtained by solving Problem~\ref{EP1}
 for the respective MRT using the method proposed in Section~III-A.
In the scenario with user transcoding, we consider one baseline scheme, i.e., Baseline 3, which
transmits message $(\mathcal{S},r_{\mathcal{S},\max})$ to all users in $\mathcal{S}$ using multicast,
where $r_{\mathcal{S},\max}\triangleq\max_{k\in\mathcal{S}}r_k,$ and uses the optimal subcarrier, power,
 and rate allocation and beamforming design as in Section~III-B.
We
evaluate the average total transmission power in the scenario without user transcoding
and the sum of the average total transmission power and transcoding power in the scenario with user
transcoding.
In the following, both measurement metrics are referred to as average power for short. We implement the proposed solutions and baseline schemes using Matlab and CVX (a Matlab software for disciplined convex programming).

In this simulation, we set $\beta_k=1$ and  $E_k=10^{-6}$ W for all $k\in\mathcal{K}$, $F_h=F_v=100^{\circ}$, $U_h\times U_v=30\times15$, $B =39$ kHz, $N=64$, $n_0 = 10^{-9}$ W, and $\alpha=1$. The elements of  $\mathbf{H}_{n,k}$, $n\in\mathcal{N}$, $k\in\mathcal{K}$, are independent and identically distributed according to $\mathcal{CN}(0,\mathbf{1}_{M\times M})$. We consider the 3DoF VR video sequence $\mathrm{Venice}$ \cite{vr-sequence1} and
use the viewing directions of 30 users for the 15-th frame
of this video sequence obtained from real measurements in \cite{vr-sequence1} as the predicted viewing directions.
To deal with possible prediction errors,  an extra $15^{\circ}$ in the four directions of the predicted FoV is transmitted for each user \cite{guo_ofdma,guo_tdma}.
The 360 VR video encoder named Kvazaar is adopted.
Set $L=5$, and choose $D_l,l\in\mathcal{L}$ as in \cite{long}.
For any $\mathcal{G}_k,k\in\mathcal{K}$, we evaluate the average power
 over 100 random realizations of small-scale channel fading coefficients.

\begin{figure}[t]
\begin{center}
 \subfigure[\small{Average power versus $K$. $M=4$, $\mathbf{r}=(2,2,3,3,4)$.}]
 {\resizebox{1.4in}{!}{\includegraphics{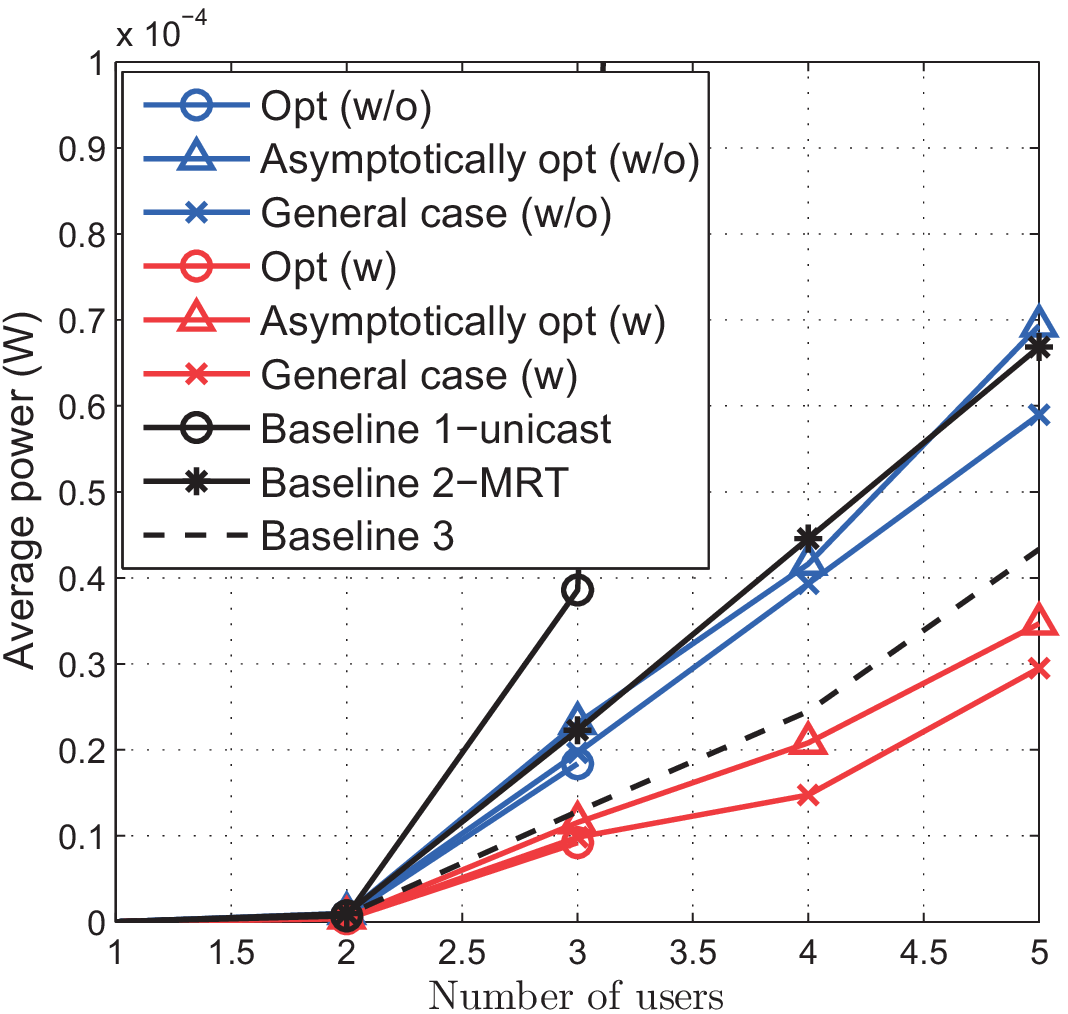}}}
  \subfigure[\small{Average power  versus $M$. $K=4$, $\mathbf{r}=(2,3,3,4)$.}]
 {\resizebox{1.4in}{!}{\includegraphics{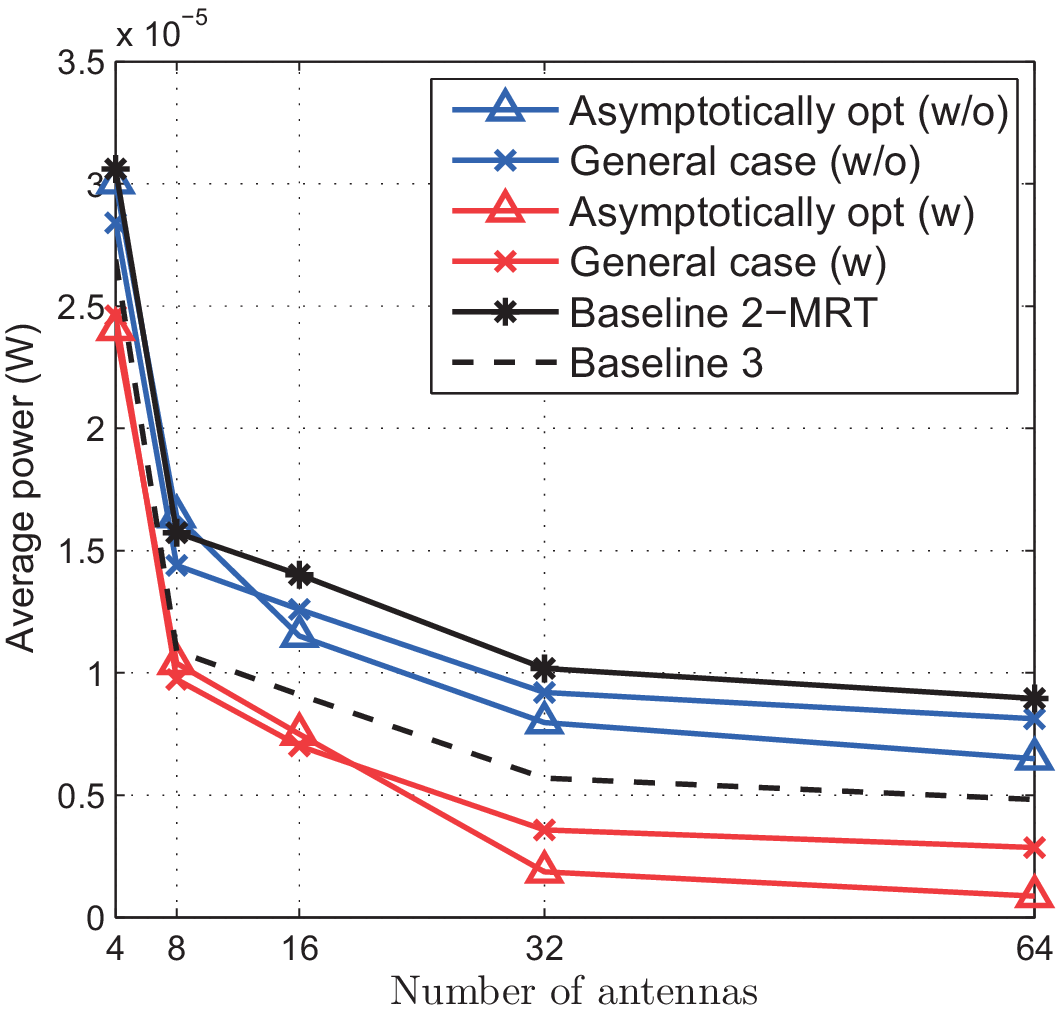}}}
 \end{center}
   \caption{\small{Average power versus $K$ and $M$. }}
   \label{fig:zipf}
\end{figure}

\begin{figure}[t]
\begin{center}
 \includegraphics[width=2in]{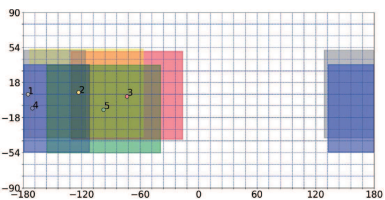}
  \end{center}
   \caption{\small{Viewing directions and corresponding FoVs of $5$ users \cite{vr-sequence1}. }}
   \label{fig:fov}
\end{figure}

First, we evaluate the average power over 1,000 random choices for
the viewing directions of 1-5 users from 30 users from \cite{vr-sequence1}.
Fig.~\ref{fig:zipf}~(a) illustrates the average power versus the number of users $K$. Since the proposed optimal solutions for small multicast groups in the scenarios without and with user transcoding are valid only for $|\mathcal{K}_{\mathcal{S},l}|\leq3, \mathcal{S}\in\mathcal{I},l\in\mathcal{L}_{\mathcal{S}}$, and $|\mathcal{S}|\leq 3,\mathcal{S}\in\mathcal{I}$, respectively. Therefore, we do not show their average powers for $K > 3$ where the abovementioned conditions are not satisfied.
We can observe that the  average powers of the proposed solutions and baseline schemes increase with $K$, as the transmission load increases with $K$.
When $K\leq3$, the
proposed solutions for the general case
achieve close-to-optimal  powers.
Given the unsatisfactory performance of Baseline~1,
 we no longer compare with it in the remaining figures.
Fig.~\ref{fig:zipf}~(b)  illustrates the average power versus the number of antennas $M$.
We can observe that the  powers achieved by the proposed solutions and baseline
schemes
decrease with $M$.  Also, when $M$ is sufficiently large, the
proposed asymptotically optimal solutions
reach close-to-optimal average powers,  demonstrating the asymptotically optimalities of the proposed solutions for the case of a large antenna array.

\begin{figure}[t]
\begin{center}
 \subfigure[\small{Average power  versus $\Delta$. $K=5$, $M=4$, $\mathbf{r}=(2,2,3,3,4)$.}]
 {\resizebox{1.4in}{!}{\includegraphics{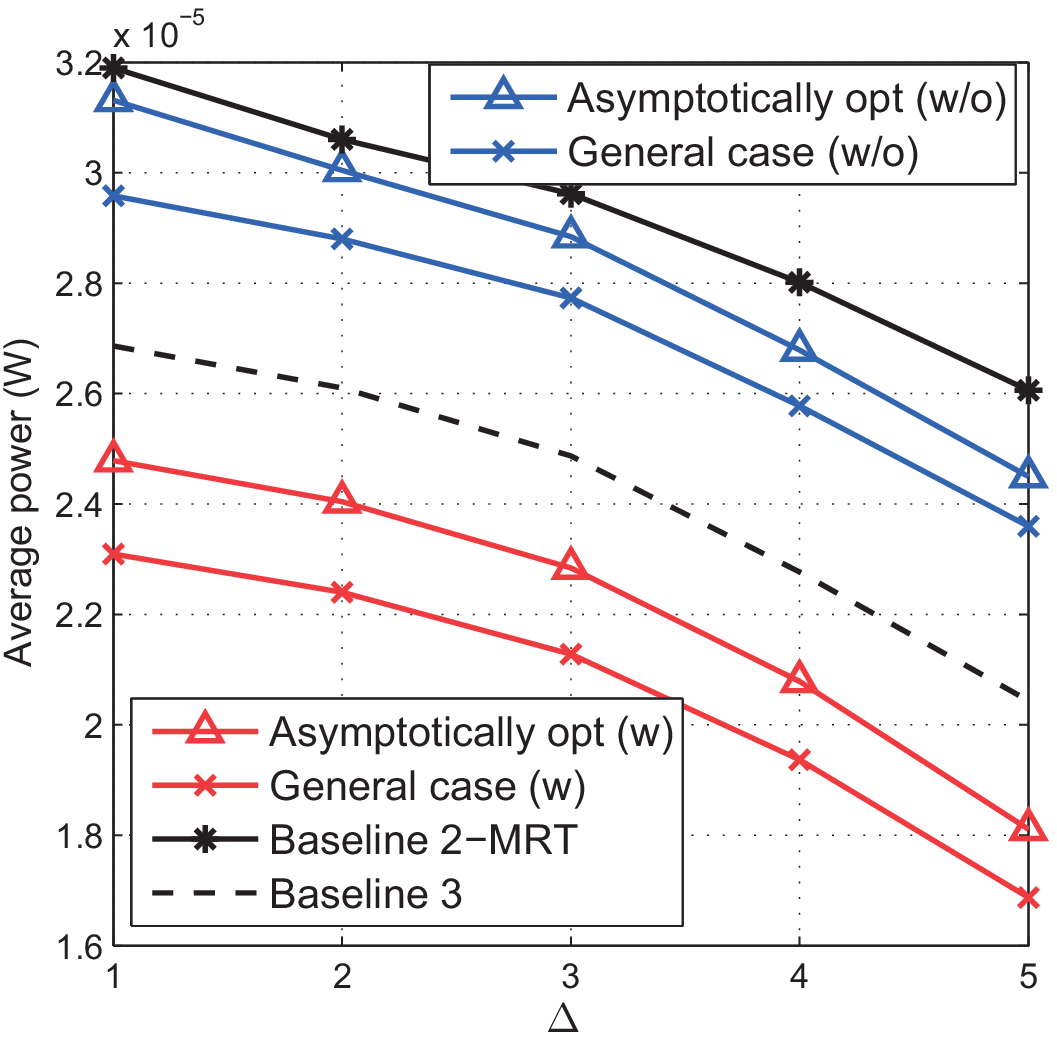}}}
 \subfigure[\small{Average power  versus $\tau$. $K=5$, $M=4$.}]
 {\resizebox{1.4in}{!}{\includegraphics{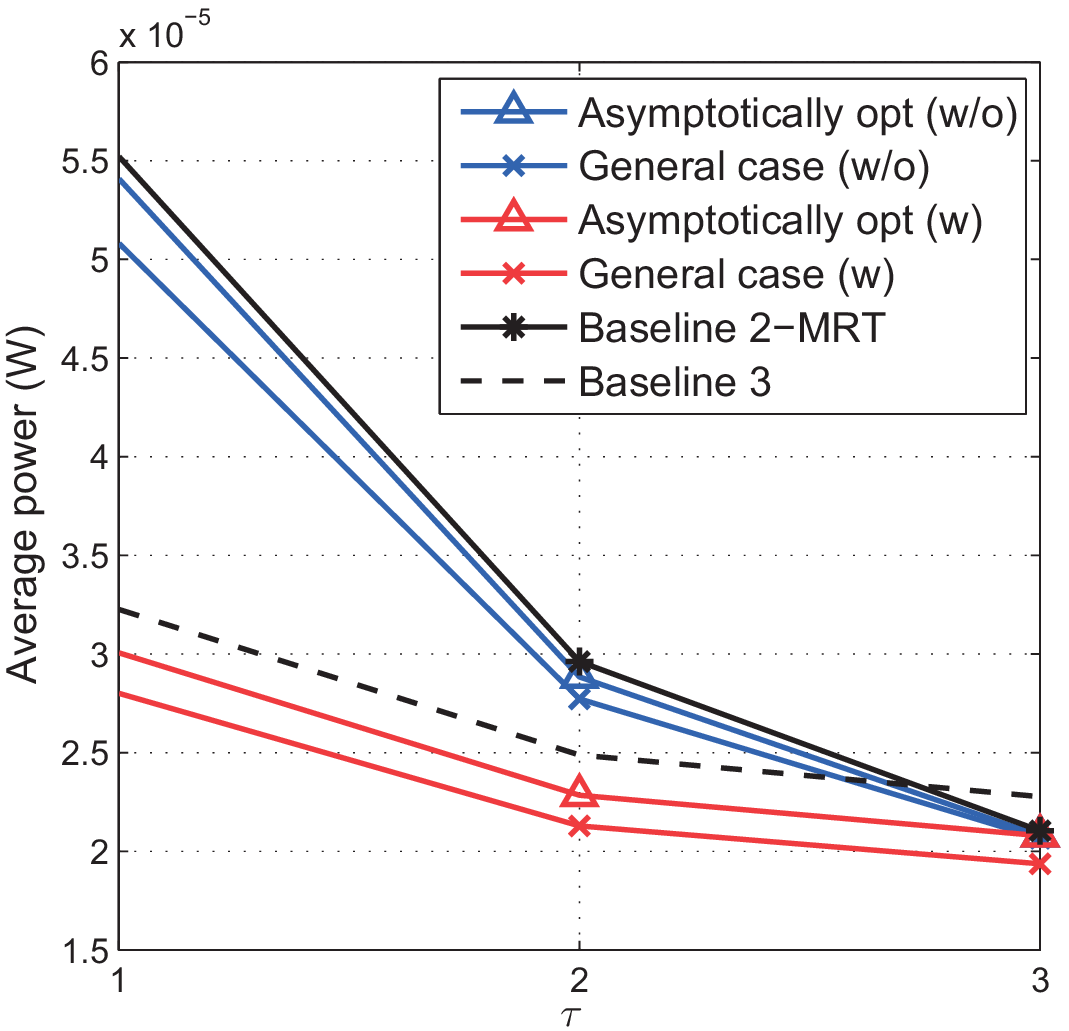}}}
 \end{center}
   \caption{\small{Average power versus $\Delta$ and $\tau$. }}
   \label{fig:fovc}
\end{figure}

Next, we show the impacts of the concentration
 of the viewing directions of all users and the similarity of the required quality levels of all users.
 We choose the viewing
 directions of 5 users out of 30 users from
 \cite{vr-sequence1}, i.e., $(\nu_k,\gamma_k)_{k\in\{1,\ldots,5\}}$,
 as shown in Fig.~\ref{fig:fov}.
 To show the impact of the concentration of the viewing directions of all users,
 based on the chosen viewing directions,
 we consider five sets of viewing directions, i.e.,
 $(\nu_1+\Delta,\gamma_1),(\nu_2+\Delta,\gamma_2),(\nu_3,\gamma_3),(\nu_4-\Delta,\gamma_4),$ and $(\nu_5-\Delta,\gamma_5)$,  $\Delta=1,\ldots,5$,
 and evaluate the corresponding average powers.
Note that $\Delta$ reflects the concentration of the viewing directions of the 5 users.
In particular, the concentration increases with $\Delta$.
Fig.~\ref{fig:fovc}~(a) shows the average power versus the concentration parameter $\Delta$. We can see that
each multicast scheme's average power
 decreases with $\Delta$, since multicast
opportunities increase with $\Delta$.
To show the impact of the similarity of the required quality levels, we consider three sets of  required quality levels, i.e.,
$\bar{\mathbf{r}}_{\tau}=(\min\{\tau,3\},\min\{\tau+1,3\},3,\max\{3,5-\tau\},\max\{3,6-\tau\})$, $\tau=1,2,3$.
Note that $\tau$ indicates the similarity of the required quality levels of the users.
Specifically, the required quality levels are closer when $\tau$ is larger.
Fig.~\ref{fig:fovc}~(b) illustrates the average power versus the similarity parameter $\tau$.
We can see that when $\tau$ increases,
the average power of each multicast scheme decreases,
due to the rise of natural multicast opportunities.
Furthermore, as $\tau$ increases, the gaps
between the average powers of the multicast schemes in the scenario without user transcoding
and those in the scenario with user transcoding
decrease, as transcoding-enabled multicast opportunities
decrease.

Fig.~\ref{fig:zipf} and Fig.~\ref{fig:fovc} show that the proposed solutions in the scenario with user transcoding outperform those in the scenario without user transcoding, which demonstrates
the importance of using transcoding-enabled multicast opportunities in reducing power
consumption.
Fig.~\ref{fig:zipf} and Fig.~\ref{fig:fovc} also show  that the
proposed solutions outperform the  baseline schemes. Specifically,
the proposed solutions in the scenario without user transcoding  outperform Baseline~1, as
 the proposed solutions utilizing multicast transmission offer   higher spectral efficiency. The proposed solutions in the scenario without user transcoding  outperform Baseline~2, as
the proposed solutions carefully choose beamforming vectors.
The  proposed solutions in the scenario with user transcoding
 outperform Baseline~3, as the proposed solutions in the scenario with user transcoding  optimally
exploit transcoding-enabled multicast opportunities to reduce power consumption.

\section{Conclusion}
This paper studied the optimal wireless streaming of a multi-quality tiled 360 VR video to multiple users in a MIMO-OFDMA system.
In the scenario without user transcoding,
we minimized  the total transmission power by optimizing the beamforming, and subcarrier, transmission power and rate allocation.
This is a challenging mixed
discrete-continuous optimization problem.
We obtained a globally optimal solution for small multicast groups, an asymptotically optimal solution for a large antenna array, and a suboptimal solution for the general case.
In the scenario with user transcoding, we minimized the weighted sum of the average total
transmission power and the transcoding power by optimizing the quality level selection, beamforming, and subcarrier, transmission power, and rate allocation. This is a more challenging two-timescale mixed discrete-continuous optimization
problem.
We obtained a globally optimal solution for small multicast groups, an asymptotically optimal solution for a large antenna array, and a low-complexity suboptimal solution for the general case.
Finally, numerical results showed that the proposed solutions have significant gains over
existing schemes.
%

\section*{Appendix A: Proof of Theorem~1}\label{app:lemma 1}

First, we obtain a problem with the same optimal value as Problem~\ref{P1}.
Let $(\bm{\mu}^{\star}(\mathbf{h}),\bm{\eta}^{\star}(\mathbf{h}),\mathbf{c}^{\star}(\mathbf{h})$, $\mathbf{w}^{\star}(\mathbf{h}))$ denote an optimal solution of Problem~\ref{P1}. By contradiction, we can easily show
\begin{align}
c^{\star}_{\mathcal{S},l,n}(\mathbf{h})=&\mu^{\star}_{\mathcal{S},l,n}(\mathbf{h})B\log_2\left(1+
\right.\nonumber\\
&\left.\frac{\eta^{\star}_{\mathcal{S},l,n}(\mathbf{h})
\min\limits_{k\in\mathcal{K}_{\mathcal{S},l}}\beta_{k}|\mathbf{h}_{k,n}^H\mathbf{w}^{\star}_{n}(\mathbf{h})|^2}
    {M\sigma^2}\right).\nonumber
\end{align}

\noindent \normalsize{Thus, equivalently, we can
 eliminate $\mathbf{c}(\mathbf{h})$ and replace the constraints in \eqref{cst:c}, \eqref{cst:minrate}, and \eqref{cst:averate} with}
\small{\begin{align}
&\sum\limits_{n\in\mathcal{N}}\mu_{\mathcal{S},l,n}(\mathbf{h})B\log_2\left(1+
    \frac{\eta_{\mathcal{S},l,n}(\mathbf{h})\min\limits_{k\in\mathcal{K}_{\mathcal{S},l}}\beta_{k}|\mathbf{h}_{k,n}^H\mathbf{w}_{n}(\mathbf{h})|^2}
    {M\sigma^2}\right)\nonumber\\
    &\geq |\mathcal{P}_{\mathcal{S}}|D_{l},~\mathcal{S}\in\mathcal{I},l\in\mathcal{L}_{\mathcal{S}}.\label{pfcst:rate}
\end{align}}

\noindent\normalsize{Define $P_{\mathcal{S},l,n}(\mathbf{h})=\mu_{\mathcal{S},l,n}(\mathbf{h})\eta_{\mathcal{S},l,n}(\mathbf{h})$, $\mathbf{\tilde{w}}_{\mathcal{S},l,n}(\mathbf{h})=\mathbf{w}_{n}(\mathbf{h})$,
and}
\begin{equation}
\widetilde{Q}_{\mathcal{S},l,n}(\mathbf{h})=\max\limits_{k\in\mathcal{K}_{\mathcal{S},l}}\frac{M\sigma^2}{\beta_{k}|\mathbf{h}_{k,n}^H\mathbf{w}_{n}(\mathbf{h})|^2},\mathcal{S}\in\mathcal{I},l\in\mathcal{L}_{\mathcal{S}},n\in\mathcal{N}.
\label{cst:Q}
\end{equation}
\normalsize{By the change of variables of $\mathbf{P}(\mathbf{h}) \triangleq (P_{\mathcal{S},l,n}(\mathbf{h}) )_{\mathcal{S}\in\mathcal{I},l\in\mathcal{L}_{\mathcal{S}},n\in\mathcal{N}}$ and $\mathbf{\tilde{w}}(\mathbf{h}) \triangleq(\tilde{\mathbf{w}}_{\mathcal{S},l,n}(\mathbf{h}) )_{\mathcal{S}\in\mathcal{I},l\in\mathcal{L}_{\mathcal{S}},n\in\mathcal{N}}$, and introducing the auxiliary variable $\mathbf{\widetilde{Q}}(\mathbf{h})\triangleq(\widetilde{Q}_{\mathcal{S},l,n}(\mathbf{h}))_{\mathcal{S}\in\mathcal{I},l\in\mathcal{L}_{\mathcal{S}},n\in\mathcal{N}}$,
the constraints in \eqref{cst:w} and \eqref{pfcst:rate} can be transformed to \eqref{cst:Q} and the following constraints:}
\begin{align}
&\|\tilde{\mathbf{w}}_{\mathcal{S},l,n}(\mathbf{h})\|_2=1,~\mathcal{S}\in\mathcal{I},l\in\mathcal{L}_{\mathcal{S}},n\in\mathcal{N},\label{pf-tw}\\
&\tilde{\mathbf{w}}_{\mathcal{S},l,n}(\mathbf{h})=\tilde{\mathbf{w}}_{\mathcal{S}',l',n}(\mathbf{h}), ~\mathcal{S},\mathcal{S}'\in\mathcal{I},l\in\mathcal{L}_{\mathcal{S}},l'\in\mathcal{L}_{\mathcal{S}'},n\in\mathcal{N},\label{pf-tw2}\\
&\sum\limits_{n\in\mathcal{N}}\mu_{\mathcal{S},l,n}(\mathbf{h})B\log_2\left(1+
    \frac{P_{\mathcal{S},l,n}(\mathbf{h})}
    {\mu_{\mathcal{S},l,n}(\mathbf{h})\widetilde{Q}_{\mathcal{S},l,n}(\mathbf{h})}\right)\nonumber\\
&\geq |\mathcal{P}_{\mathcal{S}}|D_{l},~\mathcal{S}\in\mathcal{I},l\in\mathcal{L}_{\mathcal{S}}.\label{pfcst-w0}
\end{align}

\noindent \normalsize{In addition, by contradiction, we can easily show that  \eqref{cst:Q} can be equivalently  replaced by the following constraints:}
\begin{equation}
\begin{matrix}
\widetilde{Q}_{\mathcal{S},l,n}(\mathbf{h})\geq\max\limits_{k\in\mathcal{K}_{\mathcal{S},l}}\frac{M\sigma^2}{\beta_{k}|\mathbf{h}_{k,n}^H\mathbf{\tilde{w}}_{\mathcal{S},l,n}(\mathbf{h})|^2},~\mathcal{S}\in\mathcal{I},l\in\mathcal{L}_{\mathcal{S}},n\in\mathcal{N}.
\label{cst:Q2}
\end{matrix}
\end{equation}

\noindent \normalsize{Thus,
Problem~\ref{P1} and the following problem have the same optimal value.}
\begin{Prob}[Equivalent Problem of Problem~\ref{P1}]\label{EPPP1}
\begin{align}
E^{\star}(\mathbf{h})=\min_{\bm{\mu}(\mathbf{h}),\mathbf{P}(\mathbf{h}),\mathbf{\tilde{w}}(\mathbf{h}),\mathbf{\widetilde{Q}}(\mathbf{h})}~&\frac{1}{M}\sum\limits_{n\in\mathcal{N}}\sum\limits_{\mathcal{S}\in\mathcal{I}}\sum\limits_{l\in\mathcal{L}_{\mathcal{S}}}P_{\mathcal{S},l,n}(\mathbf{h})
 \nonumber\\
    \mathrm{s.t.} ~~
    \eqref{cst:mu1},~\eqref{cst:mu2},&~\eqref{cst:p},~\eqref{pf-tw},~\eqref{pf-tw2},~\eqref{pfcst-w0},~\eqref{cst:Q2}.\nonumber
\end{align}
Denote  $(\bm{\mu}^{\star}(\mathbf{h}),\bm{P}^{\star}(\mathbf{h}),\mathbf{\tilde{w}}^{\star}(\mathbf{h}),
\widetilde{\mathbf{Q}}^{\star}(\mathbf{h}))$  an optimal solution.
\end{Prob}

Then, we obtain a problem with the same optimal value as Problem \ref{EPPP1}. Consider the following problem.
\begin{Prob}[Equivalent Problem of Problem~\ref{EPPP1}]\label{EPP1}
\begin{align}
E^{\ddagger}(\mathbf{h})=\min_{\bm{\mu}(\mathbf{h}),\mathbf{P}(\mathbf{h}),\mathbf{\tilde{w}}(\mathbf{h}),\mathbf{\widetilde{Q}}(\mathbf{h})}~&\frac{1}{M}\sum\limits_{n\in\mathcal{N}}\sum\limits_{\mathcal{S}\in\mathcal{I}}\sum\limits_{l\in\mathcal{L}_{\mathcal{S}}}P_{\mathcal{S},l,n}(\mathbf{h})
 \nonumber\\
    \mathrm{s.t.} ~~
    \eqref{cst:mu1},~&\eqref{cst:mu2},~\eqref{cst:p},~\eqref{pf-tw},~\eqref{pfcst-w0},~\eqref{cst:Q2}.\nonumber
\end{align}
\end{Prob}
Let  $(\bm{\mu}^{\ddagger}(\mathbf{h}),\bm{P}^{\ddagger}(\mathbf{h}),\mathbf{\tilde{w}}^{\ddagger}(\mathbf{h}),
\widetilde{\mathbf{Q}}^{\ddagger}(\mathbf{h}))$ denote an optimal solution.  As Problem~\ref{EPPP1} has extra constraints, i.e., \eqref{pf-tw2}, compared to Problem~\ref{EPP1}, $E^{\star}(\mathbf{h})\geq E^{\ddagger}(\mathbf{h})$. Thus, it remains to  show that $E^{\star}(\mathbf{h})\leq E^{\ddagger}(\mathbf{h})$. Based on an optimal solution of Problem~\ref{EPP1}, i.e., $(\bm{\mu}^{\ddagger}(\mathbf{h}),\bm{P}^{\ddagger}(\mathbf{h}),\mathbf{\tilde{w}}^{\ddagger}(\mathbf{h}),
\widetilde{\mathbf{Q}}^{\ddagger}(\mathbf{h}))$, we construct a feasible solution of Problem~\ref{EPPP1}, whose objective value is $E^{\ddagger}(\mathbf{h})$.
Specifically, for all $n\in\mathcal{N}$,
we construct
$\mathbf{\tilde{w}}^{\diamond}_{\mathcal{S},l,n}(\mathbf{h})=\mathbf{\tilde{w}}^{\ddagger}_{\mathcal{S}_n,l_n,n}(\mathbf{h})$,  $\mathcal{S}\in\mathcal{I}$, $l\in\mathcal{L}_{\mathcal{S}}$, where $(\mathcal{S}_n,l_n)$ satisfies $\mu_{\mathcal{S},l,n}^{\ddagger}(\mathbf{h})=1$.\footnote{Due to the constraints in \eqref{cst:mu1} and \eqref{cst:mu2}, there exists only one message $(\mathcal{S}_n,l_n)$ with $\mu_{\mathcal{S}_n,l_n,n}^{\ddagger}(\mathbf{h})=1$, for all  $n\in\mathcal{N}$.} It is obvious that
$(\bm{\mu}^{\ddagger}(\mathbf{h}),\bm{P}^{\ddagger}(\mathbf{h}),\mathbf{\tilde{w}}^{\diamond}(\mathbf{h}),
\widetilde{\mathbf{Q}}^{\ddagger}(\mathbf{h}))$ is a feasible solution of Problem~\ref{EPPP1}. Thus, we have
$E^{\star}(\mathbf{h})\leq\frac{1}{M}\sum\nolimits_{n\in\mathcal{N}}\sum\nolimits_{\mathcal{S}\in\mathcal{I}}\sum\nolimits_{l\in\mathcal{L}_{\mathcal{S}}}P^{\ddagger}_{\mathcal{S},l,n}(\mathbf{h})=E^{\ddagger}(\mathbf{h}).$
By $E^{\star}(\mathbf{h})\geq E^{\ddagger}(\mathbf{h})$ and $E^{\star}(\mathbf{h})\leq E^{\ddagger}(\mathbf{h})$, we have $E^{\star}(\mathbf{h})=E^{\ddagger}(\mathbf{h})$.

Next, we show that Problem~\ref{EPP1}
and
Problem~\ref{EP1}
 have the same optimal value. It is obvious that
Problem~\ref{EPP1} is equivalent to the following problem.
\begin{Prob}[Equivalent Problem of Problem~\ref{EPP1} for $\mathbf{h}$]\label{PFEP1}
\begin{align}
&E^{\ddagger}(\mathbf{h})=\min_{\bm{\mu}(\mathbf{h}),\mathbf{P}(\mathbf{h}) }~\frac{1}{M}\sum\limits_{n\in\mathcal{N}}\sum\limits_{\mathcal{S}\in\mathcal{I}}\sum\limits_{l\in\mathcal{L}_{\mathcal{S}}}P_{\mathcal{S},l,n}(\mathbf{h}) \nonumber\\
    &\mathrm{s.t.} ~~
    \eqref{cst:mu1},~\eqref{cst:mu2},~\eqref{cst:p},\nonumber\\
&\sum\nolimits_{n\in\mathcal{N}}\mu_{\mathcal{S},l,n}(\mathbf{h})B\log_2\left(1+
    \frac{P_{\mathcal{S},l,n}(\mathbf{h})}
    {\mu_{\mathcal{S},l,n}(\mathbf{h})\widetilde{Q}_{\mathcal{S},l,n}^{\star}(\mathbf{h})}\right)\nonumber\\
    &\geq |\mathcal{P}_{\mathcal{S}}|D_{l},~\mathcal{S}\in\mathcal{I},l\in\mathcal{L}_{\mathcal{S}},k\in\mathcal{K}_{\mathcal{S},l},\label{cst:pfq}
\end{align}
\end{Prob}
where $\widetilde{Q}_{\mathcal{S},l,n}^{\star}(\mathbf{h})$ is given by the following problem.
\begin{align}
\widetilde{Q}_{\mathcal{S},l,n}^{\star}(\mathbf{h})\triangleq
&\min_{\tilde{\mathbf{w}}_{\mathcal{S},l,n}(\mathbf{h})}\max_{k\in\mathcal{K}_{\mathcal{S},l}}\frac {M\sigma^2}{\beta_{k}|\mathbf{h}_{k,n}^H\tilde{\mathbf{w}}_{\mathcal{S},l,n}(\mathbf{h})|^2}
 \label{bf2}  \\
     &\mathrm{s.t.} ~~\eqref{pf-tw}.\nonumber
\end{align}
As
\small{\begin{align}
&Q^{\dagger}_{\mathcal{S},l,n}(\mathbf{h})\overset{(a)}=\frac{\text{tr}(\mathbf{V}_{\mathcal{S},l,n}^{\dagger}(\mathbf{h}))}
{\|\tilde{\mathbf{w}}_{\mathcal{S},l,n}^{\star}(\mathbf{h})\|_2^2}\nonumber\\
    &\overset{(b)}=\left(\min_{k\in\mathcal{K}_{\mathcal{S},l}}\frac{\text{tr}(\beta_k\mathbf{h}_{k,n} \mathbf{h}_{k,n} ^H\mathbf{V}^{\dagger}_{\mathcal{S},l,n})}{M\sigma^2}\right)\left(\max_{k\in\mathcal{K}_{\mathcal{S},l}}\frac{M\sigma^2}{\beta_{k}|\mathbf{h}_{k,n}^H\tilde{\mathbf{w}}_{\mathcal{S},l,n}^{\star}(\mathbf{h})|^2}\right)\nonumber\\
&\overset{(c)}=\max_{k\in\mathcal{K}_{\mathcal{S},l}}\frac{M\sigma^2}{\beta_{k}|\mathbf{h}_{k,n}^H\tilde{\mathbf{w}}_{\mathcal{S},l,n}^{\star}(\mathbf{h})|^2}=\tilde{Q}^{\star}_{\mathcal{S},l,n}(\mathbf{h}),\label{pf-bf1}
\end{align}}

\noindent \normalsize{where} $(a)$ is due to \eqref{pf-tw},
$(b)$ is due to Claim~2 in \cite{luo2006}, and $(c)$ is due to that
$\frac{\min_{k\in\mathcal{K}_{\mathcal{S},l}}\text{tr}(\beta_k\mathbf{h}_{k,n} \mathbf{h}_{k,n} ^H\mathbf{V}^{\dagger}_{\mathcal{S},l,n})}{M\sigma^2}=1$ (which can be easily shown by contradiction), $E^{\ddagger}(\mathbf{h})=E^{\dagger}(\mathbf{h})$.

Finally, we show that Problem \ref{P1} and Problem \ref{EP1} have the same optimal value and characterize the relation between their optimal solutions. As $E^{\star}(\mathbf{h})=E^{\ddagger}(\mathbf{h})$ and $E^{\ddagger}(\mathbf{h})=E^{\dagger}(\mathbf{h})$, we know that the optimal values of Problem \ref{P1} and Problem \ref{EP1} are identical, i.e.,
\begin{equation}
E^{\star}(\mathbf{h})=E^{\dagger}(\mathbf{h}).\label{pf-eq}
\end{equation}
In the sequel, we show that $(\bm{\mu}^{\dagger}(\mathbf{h}),\bm{\eta}^{\dagger}(\mathbf{h}),\mathbf{c}^{\dagger}(\mathbf{h}),\mathbf{w}^{\dagger}(\mathbf{h}))$  is an optimal solution of Problem~\ref{P1}.
It is obvious that $(\bm{\mu}^{\dagger}(\mathbf{h}),\bm{\eta}^{\dagger}(\mathbf{h}),\mathbf{c}^{\dagger}(\mathbf{h}),\mathbf{w}^{\dagger}(\mathbf{h}))$
satisfies the constraints in \eqref{cst:mu1}, \eqref{cst:mu2}, \eqref{cst:eta}, \eqref{cst:c}, and \eqref{cst:averate}.
It remains to show that $(\bm{\mu}^{\dagger}(\mathbf{h}),\bm{\eta}^{\dagger}(\mathbf{h}),\mathbf{c}^{\dagger}(\mathbf{h}),\mathbf{w}^{\dagger}(\mathbf{h}))$ satisfies \eqref{cst:w} and \eqref{cst:minrate}.
We have
\begin{equation}
\|\mathbf{v}_{\mathcal{S},l,n}^{\dagger}(\mathbf{h})\|_2\overset{(d)}=\sqrt{\text{tr}\left(\mathbf{V}_{\mathcal{S},l,n}^{\dagger}(\mathbf{h})\right)}=\sqrt{Q^{\dagger}_{\mathcal{S},l,n}(\mathbf{h})},\label{pf-v}
\end{equation}
where $(d)$ is due to $\mathbf{V}_{\mathcal{S},l,n}^{\dagger}(\mathbf{h})=\mathbf{v}_{\mathcal{S},l,n}^{\dagger}(\mathbf{h})(\mathbf{v}_{\mathcal{S},l,n}^{\dagger}(\mathbf{h}))^H$.
Thus, we have
\begin{equation}
\|\mathbf{w}^{\dagger}_{n}(\mathbf{h})\|_2 \overset{(e)}=\|\sum\limits_{\mathcal{S}\in\mathcal{I}}\sum\limits_{l\in\mathcal{L}_{\mathcal{S}}}\mu_{\mathcal{S},l,n}^{\dagger}(\mathbf{h})\frac{\mathbf{v}^{\dagger}_{\mathcal{S},l,n}(\mathbf{h}) }{\|\mathbf{v}_{\mathcal{S},l,n}^{\dagger}(\mathbf{h})\|_2}\|_2
\overset{(f)}=1,\label{pf-norm-w}
\end{equation}

\noindent\normalsize{where} $(e)$ is due to \eqref{thm1-w} and \eqref{pf-v}, and $(f)$ is due to \eqref{cst:mu1} and \eqref{cst:mu2}. Thus, $\mathbf{w}^{\dagger}(\mathbf{h})$ satisfies \eqref{cst:w}.
Besides, we have
\begin{align}
&\mu_{\mathcal{S},l,n}^{\dagger}(\mathbf{h})B\log_2\left(1+\frac{\eta_{\mathcal{S},l,n}^{\dagger}(\mathbf{h})\beta_{k}|\mathbf{h}_{n,k}^H\mathbf{w}_{n}(\mathbf{h})|^2}
    {M\sigma^2}\right)\nonumber\\
    &\overset{(g)}\geq\mu_{\mathcal{S},l,n}^{\dagger}(\mathbf{h})B\log_2\left(1+\frac{P_{\mathcal{S},l,n}^{\dagger}(\mathbf{h})\beta_{k}|\mathbf{h}_{n,k}^H\mathbf{v}^{\dagger}_{\mathcal{S},l,n}(\mathbf{h})|^2}
{M\sigma^2Q_{\mathcal{S},l,n}^{\dagger}(\mathbf{h})}\right)\nonumber\\
&\overset{(h)}\geq\mu_{\mathcal{S},l,n}^{\dagger}(\mathbf{h})B\log_2\left(1+\frac{P_{\mathcal{S},l,n}^{\dagger}(\mathbf{h})}
{Q_{\mathcal{S},l,n}^{\dagger}(\mathbf{h})}\right)=c_{\mathcal{S},l,n}^{\dagger}(\mathbf{h}) ,\label{pf-wc}
\end{align}
where $(g)$ is due to \eqref{thm1-eta}, \eqref{thm1-w},  \eqref{cst:mu1} and \eqref{cst:mu2}, and $(h)$ is due to  \eqref{cst:bf1} and $\mathbf{V}_{\mathcal{S},l,n}^{\dagger}(\mathbf{h})=\mathbf{v}_{\mathcal{S},l,n}^{\dagger}(\mathbf{h})(\mathbf{v}_{\mathcal{S},l,n}^{\dagger}(\mathbf{h}))^H$.
Thus, $(\bm{\mu}^{\dagger}(\mathbf{h}),\bm{\eta}^{\dagger}(\mathbf{h}),\mathbf{c}^{\dagger}(\mathbf{h}),\mathbf{w}^{\dagger}(\mathbf{h}))$ satisfies \eqref{cst:minrate}. Therefore, $(\bm{\mu}^{\dagger}(\mathbf{h}),\bm{\eta}^{\dagger}(\mathbf{h}),\mathbf{c}^{\dagger}(\mathbf{h}),\mathbf{w}^{\dagger}(\mathbf{h}))$ is a feasible solution of Problem~\ref{P1}, implying
\begin{align}
E^{\star}(\mathbf{h})&\leq\sum\nolimits_{n\in\mathcal{N}}\sum\nolimits_{\mathcal{S}\in\mathcal{I}}\sum\nolimits_{l\in\mathcal{L}_{\mathcal{S}}}
\frac{\mu^{\dagger}_{\mathcal{S},l,n}(\mathbf{h}) \eta^{\dagger}_{\mathcal{S},l,n}(\mathbf{h})}{M}\nonumber\\
&\overset{(i)}\leq\sum\limits_{n\in\mathcal{N}}\sum\limits_{\mathcal{S}\in\mathcal{I}}\sum\limits_{l\in\mathcal{L}_{\mathcal{S}}} \frac{P_{\mathcal{S},l,n}^{\dagger}(\mathbf{h})}{M}
=E^{\dagger}(\mathbf{h})\overset{(j)}=E^{\star}(\mathbf{h}),
\end{align}
where $(i)$ is due to  \eqref{cst:mu1}, \eqref{cst:mu2} and \eqref{thm1-eta}, and $(j)$ is due to \eqref{pf-eq}.
Thus, we can conclude that $(\bm{\mu}^{\dagger}(\mathbf{h}),\bm{\eta}^{\dagger}(\mathbf{h}),\mathbf{c}^{\dagger}(\mathbf{h}),\mathbf{w}^{\dagger}(\mathbf{h}))$ achieves the optimal value of Problem~\ref{P1} and is an optimal  solution of Problem~\ref{P1}.

\section*{Appendix B: Proof of Theorem 2}\label{app:lemma 1}

First,  we show that the problem in \eqref{bf2} and Problem~\ref{BF1} are equivalent.
We have
\begin{align}
\tilde{\mathbf{w}}_{\mathcal{S},l,n}^{\star}(\mathbf{h})\overset{(a)}=&\frac{\tilde{\mathbf{w}}_{\mathcal{S},l,n}^{\star}(\mathbf{h})}
{\|\tilde{\mathbf{w}}_{\mathcal{S},l,n}^{\star}(\mathbf{h})\|}\overset{(b)}=
\frac{\mathbf{v}_{\mathcal{S},l,n}^{\dagger}(\mathbf{h})}
{\|\mathbf{v}_{\mathcal{S},l,n}^{\dagger}(\mathbf{h})\|}\overset{(c)}=\frac{\mathbf{v}_{\mathcal{S},l,n}^{\dagger}(\mathbf{h})}
{\sqrt{Q^{\dagger}_{\mathcal{S},l,n}(\mathbf{h})}}\nonumber\\
\overset{(d)}=&\frac{\mathbf{v}_{\mathcal{S},l,n}^{\dagger}(\mathbf{h})}
{\sqrt{\max_{k\in\mathcal{K}_{\mathcal{S},l}}\frac{M\sigma^2}{\beta_{k}|\mathbf{h}_{k,n}^H\tilde{\mathbf{w}}_{\mathcal{S},l,n}^{\star}(\mathbf{h})|^2}}},\label{pf-bf2}
\end{align}
where $(a)$ is due to  \eqref{pf-tw},
$(b)$ is due to Claim~2 in \cite{luo2006},  $(c)$ is due to \eqref{pf-v}, and $(d)$ is due to \eqref{pf-bf1}.
Thus, we have
\begin{align}
&\mathbf{V}_{\mathcal{S},l,n}^{\dagger}(\mathbf{h})=\mathbf{v}_{\mathcal{S},l,n}^{\dagger}(\mathbf{h})(\mathbf{v}_{\mathcal{S},l,n}^{\dagger}(\mathbf{h}))^H
\nonumber\\&=\tilde{\mathbf{w}}_{\mathcal{S},l,n}^{\star}(\mathbf{h})(\tilde{\mathbf{w}}_{\mathcal{S},l,n}^{\star}(\mathbf{h}))^H\max_{k\in\mathcal{K}_{\mathcal{S},l}}\frac{M\sigma^2}{\beta_{k}|\mathbf{h}_{k,n}^H\tilde{\mathbf{w}}_{\mathcal{S},l,n}^{\star}(\mathbf{h})|^2}.
\label{pf-eq2}
\end{align}
By \eqref{pf-bf1} and \eqref{pf-eq2}, we can conclude that  the problem in \eqref{bf2} and Problem~\ref{BF1} are equivalent.

Next, we obtain an asymptotically optimal solution of the problem in \eqref{bf2}.
 Following the proof of Theorem~1 in
\cite{xiang},
we can show that $\mathbf{\tilde{w}}_{\mathcal{S},l,n}^{\ast}(\mathbf{h})=\frac{\sum\nolimits_{k\in\mathcal{K}_{\mathcal{S},l}}\xi_{n,k}^{\ast}\mathbf{h}_{n,k}}{\left\|\sum\nolimits_{k\in\mathcal{K}_{\mathcal{S},l}}\xi_{n,k}^{\ast}\mathbf{h}_{n,k}\right\|_2}$ is asymptotically optimal for the problem in  \eqref{bf2},
where $\xi_{n,k}^{\ast}$
is an optimal solution of the following problem:
\begin{equation}
\min_{\xi_{n,k}}\max_{k\in\mathcal{K}_{\mathcal{S},l}}\frac{\sigma^2\sum\nolimits_{j\in\mathcal{K}}\xi_{n,j}^2}
{\beta_k\xi_{n,k}^2}.\label{bf3}
\end{equation}
This problem is similar to Problem~$\mathcal{Q}$ in \cite{xiang}. Using  the method  proposed in \cite{xiang},
 we have
$\xi^{\ast}_{n,k}=\frac{1}{\sqrt{\beta_k}}$.
Thus, the asymptotically optimal solution of the problem in \eqref{bf2} can be written as $\mathbf{\tilde{w}}_{\mathcal{S},l,n}^{\ast}(\mathbf{h})=
\frac{\sum\nolimits_{k\in\mathcal{K}_{\mathcal{S},l}}\frac{1}{\sqrt{\beta_k}}\mathbf{h}_{n,k}}{\left\|\sum\nolimits_{k\in\mathcal{K}_{\mathcal{S},l}}\frac{1}{\sqrt{\beta_k}}\mathbf{h}_{n,k}\right\|_2}$.

Finally, we show that $\mathbf{V}_{\mathcal{S},l,n}^{\ast}(\mathbf{h})$ is an asymptotically optimal solution of Problem~\ref{BF1}.
By $\mathbf{V}_{\mathcal{S},l,n}^{\ast}(\mathbf{h})=\mathbf{v}_{\mathcal{S},l,n}^{\ast}(\mathbf{h})(\mathbf{v}_{\mathcal{S},l,n}^{\ast}(\mathbf{h}))^H$ and \eqref{thm2-v}, we have
\small{\begin{equation}
\mathbf{V}_{\mathcal{S},l,n}^{\ast}(\mathbf{h})=\tilde{\mathbf{w}}_{\mathcal{S},l,n}^{\ast}(\mathbf{h})(\tilde{\mathbf{w}}_{\mathcal{S},l,n}^{\ast}(\mathbf{h}))^H\max_{k\in\mathcal{K}_{\mathcal{S},l}}\frac{M\sigma^2}{\beta_{k}|\mathbf{h}_{k,n}^H\tilde{\mathbf{w}}_{\mathcal{S},l,n}^{\ast}(\mathbf{h})|^2}.
\label{pf-eq3}
\end{equation}}

\noindent\noindent \normalsize{Since} $\tilde{\mathbf{w}}_{\mathcal{S},l,n}^{\ast}(\mathbf{h})$ is an asymptotically optimal solution of the problem in \eqref{bf2} , by \eqref{pf-eq2} and \eqref{pf-eq3}, we can conclude that $\mathbf{V}_{\mathcal{S},l,n}^{\ast}(\mathbf{h})$ is an asymptotically optimal solution of Problem~\ref{BF1}.

\section*{Appendix C: Proof of Theorem 3}\label{app:lemma 1}
First, we show that the optimal value of Problem \ref{AAP2} is no greater than that of Problem \ref{EAP2}, i.e., $E_t^{\dagger}\leq \bar{E}_t^{\ast}$.
Based on an optimal solution of Problem~\ref{EAP2}, i.e., $(\bar{\mathbf{N}}^{\ast},\bar{\mathbf{P}}^{\ast},\mathbf{x}^{\ast})$, we construct a feasible
solution of Problem~\ref{AAP2}, whose objective value is $\bar{E}^{\ast}$.
Specifically, we construct $\bar{\mu}^{\ast}_{\mathcal{S},l,n}=\frac{\bar{N}^{\ast}_{\mathcal{S},l}}{N}$,
$\bar{p}^{\ast}_{\mathcal{S},l,n}=\frac{\bar{P}^{\ast}_{\mathcal{S},l}}{N}$, $\mathcal{S}\in\mathcal{I},l\in\mathcal{L},n\in\mathcal{N}$.
Then, it is obvious that $(\bm{\bar{\mu}}^{\ast},\bm{\bar{p}}^{\ast},\mathbf{x}^{\ast})$ satisfies the constraints  of Problem~\ref{AAP2}, implying that it  is a feasible solution of Problem~\ref{AAP2}.
Besides,  we have
\begin{equation}
E_{t}^{\dagger}\leq\sum\limits_{n\in\mathcal{N}}\sum\limits_{\mathcal{S}\in\mathcal{I}}\sum\limits_{l\in\mathcal{L}}\frac{\bar{p}^{\ast}_{\mathcal{S},l,n}}{M}
+\alpha E_{\text{tc}}(\mathbf{x}^{\ast})\overset{(a)}=\bar{E}_t^{\ast},
 \label{pf3-e1}
\end{equation}
where $(a)$ is due to  $\bar{\mu}^{\ast}_{\mathcal{S},l,n}=\frac{\bar{N}^{\ast}_{\mathcal{S},l}}{N}$,
$\bar{p}^{\ast}_{\mathcal{S},l,n}=\frac{\bar{P}^{\ast}_{\mathcal{S},l}}{N}$.

Next, we show that the optimal value of Problem \ref{AAP2} is no smaller than that of Problem \ref{EAP2}, i.e., $E_t^{\dagger}\geq \bar{E}_t^{\ast}$.
 Base on an optimal solution of Problem~\ref{AAP2}, i.e., $(\bar{\bm{\mu}}^{\dagger},\bar{\bm{p}}^{\dagger},\mathbf{x}^{\dagger})$,
we construct a feasible solution of Problem~\ref{EAP2}, whose  objective value is $E_t^{\dagger}$.
Specifically, we construct $\bar{N}^{\dagger}_{\mathcal{S},l}=\sum\nolimits_{n\in\mathcal{N}}\bar{\mu}^{\dagger}_{\mathcal{S},l,n}$,
$\bar{P}^{\dagger}_{\mathcal{S},l}=\sum\nolimits_{n\in\mathcal{N}}\bar{p}^{\dagger}_{\mathcal{S},l,n}$, $\mathcal{S}\in\mathcal{I},l\in\mathcal{L},n\in\mathcal{N}$.
 It is obvious that $(\bm{\bar{N}}^{\dagger},\bm{\bar{P}}^{\dagger},\mathbf{x}^{\dagger})$ satisfies the constraints in \eqref{cst:x},~\eqref{cst:sumx},~\eqref{cst:sumlx}, \eqref{cst:ap2-N}, \eqref{cst:ap2-P}, and \eqref{cst:ap2-sumN}.
We also have
\begin{align}
 &\frac{1}{N}\bar{N}^{\dagger}_{\mathcal{S},l}B\log_2\left(1+
    \frac{\bar{P}^{\dagger}_{\mathcal{S},l}}
    {\bar{Q}_{k}\bar{N}^{\dagger}_{\mathcal{S},l}}\right)\nonumber\\
& =\left(\sum\limits_{n\in\mathcal{N}}\frac{1}{N}\bar{\mu}^{\dagger}_{\mathcal{S},l,n}\right)B\log_2\left(1+
    \frac{\sum\limits_{n\in\mathcal{N}}\frac{1}{N}\bar{p}^{\dagger}_{\mathcal{S},l,n}}
    {\bar{Q}_{k}\sum\limits_{n\in\mathcal{N}}\frac{1}{N}\bar{\mu}_{\mathcal{S},l,n}}\right)\nonumber\\
 &\overset{(b)}\geq\sum\limits_{n\in\mathcal{N}}\bar{\mu}^{\dagger}_{\mathcal{S},l,n}B\log_2\left(1+
    \frac{\bar{p}^{\dagger}_{\mathcal{S},l,n}}
    {\bar{Q}_{k}\bar{\mu}^{\dagger}_{\mathcal{S},l,n}}\right)\nonumber\\
 &\overset{(c)}\geq |\mathcal{P}_{\mathcal{S}}|D_lx_{\mathcal{S},l,k},~\mathcal{S}\in\mathcal{I},l\in\mathcal{L},k\in\mathcal{S},n\in\mathcal{N},\label{pf-rate3}
\end{align}

\noindent\normalsize{where} $(b)$
is due to the concavity of  $\bar{\mu}_{\mathcal{S},l,n}B\log_2\left(1+
    \frac{\bar{p}_{\mathcal{S},l,n}}
    {\bar{Q}_{k}\bar{\mu}_{\mathcal{S},l,n}}\right)$   in $(\bar{\mu}_{\mathcal{S},l,n},\bar{p}_{\mathcal{S},l,n})$, and $(c)$ is due to \eqref{cst:minrate2}. By  \eqref{pf-rate3}, we know that  $(\bm{\bar{N}}^{\dagger},\bm{\bar{P}}^{\dagger},\mathbf{x}^{\dagger})$ satisfies the constraints in \eqref{cst:ap2-rate}.
Thus, $(\bm{\bar{N}}^{\dagger},\bm{\bar{P}}^{\dagger},\mathbf{x}^{\dagger})$ is a feasible solution of Problem~\ref{EAP2}.
In addition,  we have
\begin{equation}
E_t^{\dagger}\overset{(d)}=\frac{1}{M}\sum\nolimits_{\mathcal{S}\in\mathcal{I}}\sum\nolimits_{l\in\mathcal{L}}\bar{P}^{\dagger}_{\mathcal{S},l}
+\alpha E_{\text{tc}}(\mathbf{x}^{\dagger})\geq\bar{E}_t^{\ast},
 \label{pf3-e2}
\end{equation}
where $(d)$ is due to $\bar{N}^{\dagger}_{\mathcal{S},l}=\sum\nolimits_{n\in\mathcal{N}}\bar{\mu}^{\dagger}_{\mathcal{S},l,n}$,
$\bar{P}^{\dagger}_{\mathcal{S},l}=\sum\nolimits_{n\in\mathcal{N}}\bar{p}^{\dagger}_{\mathcal{S},l,n}$, $\mathcal{S}\in\mathcal{I},l\in\mathcal{L},n\in\mathcal{N}$.

Finally, we show that the optimal values of Problem \ref{AAP2} and Problem \ref{EAP2} are identical and characterize the relatioship between their optimal solutions. By \eqref{pf3-e1} and \eqref{pf3-e2},
we have
$E_t^{\dagger}=\bar{E}_t^{\ast}$.
By $E_t^{\dagger}=\bar{E}_t^{\ast}$ and \eqref{pf3-e1}, we know that
$(\bm{\bar{\mu}}^{\ast},\bm{\bar{p}}^{\ast},\mathbf{x}^{\ast})$ is an optimal solution of Problem~\ref{AAP2}.
By $E_t^{\dagger}=\bar{E}_t^{\ast}$ and \eqref{pf3-e2}, we know that
$(\bm{\bar{N}}^{\dagger},\bm{\bar{P}}^{\dagger},\mathbf{x}^{\dagger})$  is an optimal solution of Problem~\ref{EAP2}.
Therefore, the proof of Theorem~3 is completed.

\bibliographystyle{IEEEtran}

\bibliographystyle{IEEEtran}

\end{document}